\documentclass[12pt,a4paper]{article}
\usepackage[myheadings]{fullpage}
\usepackage{lmodern,amsmath,amssymb}
\usepackage{fancyhdr}
\usepackage{mathrsfs}
\usepackage{apacite}
\numberwithin{equation}{section}
\usepackage{eqnarray,amsmath}
\usepackage{lastpage}
\usepackage{graphicx, wrapfig, subcaption, setspace, booktabs}
\usepackage[font=small, labelfont=bf]{caption}
\usepackage{fourier}
\usepackage{etoolbox}
\usepackage[protrusion=true, expansion=true]{microtype}
\usepackage[english]{babel}
\usepackage{url, lipsum}

\providecommand{\keywords}[1]
{
  \small	
  \textbf{\textit{Keywords---}} #1
}

\newcommand{\beq} {\begin{equation}}
\newcommand{\enq} {\end{equation}}
\newcommand{\ber} {\begin {eqnarray}}
\newcommand{\enr} {\end {eqnarray}}

\newcommand {\er}[1] {equation (\ref{#1}) }
\newcommand {\ern}[1] {equation (\ref{#1})}

\newcommand {\Er}[1] {Equation (\ref{#1}) }

\title{Generalized Cross Helicity in Non-ideal
Magnetohydrodynamics}
\author{Prachi Sharma and Asher Yahalom{\thanks{\textsuperscript{email address for correspondence: asya@ariel.ac.il}}}   \\
        \small Ariel University, Kiryat Hamada POB 3, Ariel 40700, Israel \\
}
\date{}
\begin{document}
\maketitle
\begin{abstract}
    The objective of the present paper is to investigate the constancy of the topological invariant denoted non-barotropic generalized cross helicity in the case of non-ideal  magnetohydrodynamic (MHD). Existing work considers only ideal barotropic MHD and ideal non-barotropic MHD. The non-ideal MHD case was not explored probably because of its mathematical complexity. Here we consider dissipative processes in the form of thermal conduction, finite electrical conductivity and viscosity and the effect of these processes on the cross helicity conservation. Analytical approach has been adopted to obtain the mathematical expressions for the time derivative of cross helicity. Obtained results show, that the generalized cross helicity is not conserved in the non-ideal MHD limit and indicate which processes affect the helicity and which do not. Furthermore, we indicate the configurations in which this topological constant is conserved despite the dissipative processes.
\end{abstract}

\section{Introduction}

Topological invariants have always been useful for several decades, and there are such invariants in MHD flows. For example, the importance of two helicities i.e., magnetic helicity and cross-helicity have long been discussed in relation to the controlled nuclear fusion problem and in numerous astrophysical scenarios. Earlier work \cite{yahalom2008simplified,yahalom2013aharonov,yahalom1995helicity} have studied the relations between the helicities and symmetries of ideal MHD. MHD connects Maxwell's equations with hydrodynamics of highly conductive flows to explain the macroscopic behaviour of conducting fluid such as plasma. However ideal MHD does not describe precisely the behaviour of real plasmas and this is the main motivation to study non-ideal MHD. Some important realistic processes are missing in the ideal description such as resistive heating, heat conduction and viscous effects. Viscosity plays an important role on dissipation scale while investigating the plasma turbulence in solar wind and elsewhere. Similarly magnetic diffusivity is one of the reasons for the magnetic reconnection phenomena. Thermal conductivity is also a substantial process to study the real picture of plasmas. It causes the perturbations of physical variables spread out through a plasmas. These essential properties of all three dissipative processes are the stimulus for the authors to make this current analysis.

The mathematical expression for cross helicity (correlation between the velocity and magnetic-field) is given by
 \cite{woltjer1958theorem,woltjer1958hydromagnetic}:
\begin{equation}
    H_C=\int\Vec{B}\cdot\Vec{v}\hspace{0.1cm}d^3 x
\label{CH}
\end{equation}
in which the integral is taken over the entire flow domain. Here $H_C$ is conserved for barotropic or incompressible MHD (but not for non-barotropic MHD) and is given a topological interpretation in terms of the knottiness of magnetic and flow field lines.
A generalization for non-barotropic MHD of this quantity was given by \cite{webb2014localAA,webb2014localBB}. This resembles the generalization
of barotropic fluid dynamics conserved quantities including helicity to non barotropic flows including topological constants of motion derived by \cite{mobbs1981some}.
Conservation law of cross helicity for non barotropic MHD has been discussed by \cite{webb2015multi} in a multi-symplectic formulation of MHD. A potential vorticity conservation equation for non-barotropic MHD was derived by \cite{webb2015potential} by using Noether’s second theorem.

Recently the non-barotropic cross-helicity was generalized using additional label translation symmetry groups ($\chi$ and $\eta$ translations) \cite{yahalom2019new}, this led to additional topological conservation laws, the $\chi$ and $\eta$ cross-helicities. The functions $\chi$ and $\eta$ are sometimes denoted ‘Euler potentials’, ‘Clebsch variables’ and also ‘flux representation functions’ \cite{hazeltine2003plasma}.

Cross helicity is expected to play an important role in several MHD plasma phenomena such as global magnetic-field generation, turbulence suppression, etc. It provides a measure of the degree of linkage of the vortex tubes of the velocity field with the flux tubes of the magnetic field. Cross helicity plays an important role in turbulent dynamo \cite{yokoi2013cross}.
The cross helicity density conservation law for barotropic flows is important in MHD turbulence theory \cite{zhou1990transport,zhou1990models,zank2011transport}. \cite{verma2004statistical} has discussed MHD turbulence in his review paper in detail. He has examined the Alfv\'{e}nic MHD turbulence with zero and non-zero helicities. Plasma velocity and magnetic field measurements from the Voyager 2 mission are used to study solar wind turbulence in the slow solar wind \cite{iovieno2016cross} and characterize its cross helicity. The energy fluxes of MHD turbulence provide a measure for transfers of energy among velocity and magnetic fields \cite{verma2021variable,verma2019energy}.

Magnetic helicity characterizes the topological features of magnetic field lines\cite{woltjer1958theorem,moffatt1969degree,moffatt1978field,webb2014localAA,
webb2014localBB,moffatt1995helicity}. Numerous visible features of magnetic field structures can be computed by magnetic helicity. The integral of magnetic helicity is developed by \cite{berger1984topological,finn1985magnetic}. In other words, it can be described as flux surface quantity and believe to be conserved in ideal MHD. \cite{faraco2020proof} has already shown the  conservation of magnetic helicity in turbulent flows. However when flux tubes diffuse through one another on resistive time scales, magnetic helicity dissipates\cite{barnes1986experimental}.\cite{webb2017magnetohydrodynamic} has attempted to understand the role of gauge symmetry responsible for the conservation of magnetic helicity. \cite{candelaresi2021stability} has studied the role of magnetic helicity in plasma stabilization in detail by doing series of experiments and numerical simulations. Further, its important role in determining the structures, dynamics and heating of the solar corona has been well explained by \cite{knizhnik2019role}.

Here we give a complete mathematical analysis for generalized cross helicity conservation and show this is affected by non ideal processes.

The structure of this paper is as follows:
The second section deals with the basic equations for non-ideal non-barotropic MHD. In the section that follows we introduce the modified entropy equation for the present case and finally cross helicity conservation is discussed in the last part.

\section{Standard Formulation of Non-ideal Non-barotropic MHD}

The standard set of equations solved for non-ideal non-barotropic MHD is given below (Here we use EMU system of units):
\beq
 \rho  \frac{d\vec{v}}{d t}  =  \rho \left[\frac {\partial \vec{v}}{\partial t} + (\vec{v} \cdot\vec{\nabla})\vec{v}\right]
   =  - \vec{\nabla} p + \vec J \times \vec{B}-\rho \vec{\nabla}\phi +\frac{\partial \sigma_{ik}^\prime}{\partial x_k}
   =  - \vec{\nabla} p +\frac {(\vec{\nabla} \times \vec{B})\times \vec{B}}{4\pi}-\rho \vec{\nabla}\phi +\frac{\partial \sigma_{ik}^\prime}{\partial x_k}
\label{eqofmo}
\enq
\begin{equation}
    \frac {\partial \rho}{\partial t} + \vec{\nabla} \cdot (\rho \vec{v}) =  0
\label{cont}
\end{equation}
\begin{equation}
    \vec{\nabla}\cdot\vec{B}=0
\label{magfl}
\end{equation}
\begin{equation}
    \frac {\partial \vec{B}}{\partial t}=\vec{\nabla} \times (\vec{v}\times\ \vec{B}) +\frac{\eta}{4\pi}\nabla^2\vec{B}
\label{Bdiff}
\end{equation}
The following notations are utilized: $\frac{\partial}{\partial t}$ is the partial temporal derivative. $\frac{d}{d t}=\frac{\partial}{\partial t}+\vec{v}\cdot\vec{\nabla}$ is the temporal material derivative or Lagrangian time derivative. $\vec \nabla$ has its standard meaning in vector calculus.  $\rho$ is the fluid density,$\vec{v}$ is the velocity of fluid, $\phi$ is a gravitational potential and $p$ is the pressure which depends through the equation of state on the density and entropy (the non-barotropic case).
The stress tensor is defined as:
\beq
\sigma_{ik}^\prime=\mu(\frac{\partial v_i}{\partial x_k}+\frac{\partial v_k}{\partial x_i}-\frac{2}{3}\delta_{ik}\frac{\partial v_l}{\partial x_l})
\label{visten}
\enq
in which $\mu$ is a coefficient of kinematic viscosity. Notice that we take coefficient of second viscosity (or volume viscosity) is zero for the sake of simplicity. According to classical kinetic theory, viscosity arises from collisions between particles. The current density $\vec J$ and
the magnetic field are related by Ampere's law:
\beq
\vec{\nabla}\times\vec{B}=4\pi\vec{J}
\label{ampere}
\enq
Note that Maxwell’s displacement current is often be neglected due to its smallness in MHD dynamics in non-relativistic regime. The magnetic diffusivity $\eta$ originates from Ohm's law:
\beq
\vec{E}+\vec{v}\times\vec{B}=\eta\vec{J}
\label{ohmslaw}
\enq
of non-ideal MHD, where $\vec E$ is the electric field. Ohm's law is another manifestation of collisions in the plasma. Combining Ohm's equation with Faraday's
equation:
\beq
\frac {\partial \vec{B}}{\partial t}= - \vec \nabla \times  \vec{E}
\label{Fareq}
\enq
and Ampere's law will yield \ern{Bdiff}. In this form $\eta$ resembles a diffusion coefficient describing the diffusion of the magnetic field through a conducting medium of finite conductivity.
The justification for those equations and the conditions under which they apply can be found in standard books on MHD (see for example \cite{batchelor1967introduction,sturrock1994plasma,LANDAU1987192,kundu2015fluid,ogilvie2016astrophysical}).
Let us now introduce the energy equation in the case  of non-ideal MHD.

\subsection{Energy Equation for non-ideal MHD}

we construct the total energy equation piece by piece. Similar formalism has been adopted by \cite{ogilvie2016astrophysical} in ideal MHD.

\subsubsection{Kinetic Energy}
\begin{equation}
    \rho\frac{d}{dt}(\frac{1}{2}v^2)=\rho\vec{v}\cdot\frac{d\vec{v}}{dt}=-\rho\vec{v}\cdot\vec{\nabla}\phi-\vec{v}\cdot\vec{\nabla}p+\frac{1}{4\pi}\vec{v}\cdot[(\vec{\nabla}\times\vec{B})\times\vec{B}]+{v_i}\cdot\frac{\partial \sigma_{ik}^\prime}{\partial x_k}
\label{kinen}
\end{equation}
substituted the value of $\frac{d\vec{v}}{dt}$ from momentum equation $(2.1)$.

\subsubsection{Potential energy}

Let us assume that $\phi$ has no explicit dependence on $t$:
\begin{equation}
    \rho\frac{d\phi}{dt}=\rho\vec{v}\cdot\vec{\nabla}\phi
\label{grnen}
\end{equation}
Here $\phi$ is a potential which can be due to gravity or electric conservative forces.

\subsubsection{Internal (thermal) energy}

Using the fundamental thermodynamic identity for the specific energy density: $d\epsilon=Tds-pdV$
or $d\epsilon=Tds+\frac{p}{\rho^2}d\rho$ as per unit mass the specific volume is $V = \frac{1}{\rho}$. Here $s$ is entropy per unit mass or specific entropy, $T$ is temperature.
After taking the time derivative and multiplying with the $\rho$, we obtain:
\beq
 \rho\frac{d\epsilon}{dt}=\rho T\frac{d s}{dt}+\frac{p}{\rho}\frac{d\rho}{dt}
 \label{intnen}
\enq
 Now substitute the value of $\frac{d\rho}{dt}$ from continuity equation \ern{cont} (i.e., $\frac{d\rho}{dt}+\rho\vec{\nabla}\cdot\vec{v}=0$)
\begin{equation}
    \rho\frac{d\epsilon}{dt}=\rho T\frac{d s}{dt}-p\vec{\nabla}\cdot\vec{v}
    \label{intnen2}
\end{equation}
Next, if the temperature of the fluid is not constant throughout its volume, there will be a transfer of heat known as thermal conduction. Thermal conduction heat flux density $\vec q$ \cite{LANDAU1987192} is given by:
\begin{equation}
   \vec q=-k \vec{\nabla}T
\label{qheatflux}
\end{equation}
where $k$ is thermal conductivity. Heat flow can be represented by adding and subtracting the divergence of a heat flux to the energy equation.
So, the sum of these three terms:
\begin{equation}
    \rho\frac{d}{dt}(\frac{1}{2}v^2+\phi+\epsilon)=-\vec{\nabla}\cdot(p\vec{v})+\rho T\frac{d s}{dt}+{v_i}\cdot\frac{\partial \sigma_{ik}^\prime}{\partial x_k}+\vec{\nabla}\cdot(k\vec{\nabla}T)+\frac{1}{4\pi}\vec{v}\cdot[(\vec{\nabla}\times\vec{B})\times\vec{B}]-\vec{\nabla}\cdot(k\vec{\nabla}T)
\end{equation}
The second last term can be rewritten using \ern{ohmslaw} as:
\begin{equation}
    \frac{1}{4\pi}\vec{v}\cdot[(\vec{\nabla}\times\vec{B})\times\vec{B}]=
    \frac{1}{4\pi}(\vec{\nabla}\times\vec{B})\cdot(-\vec{v}\times\vec{B})=
    \frac{1}{4\pi}(\vec{\nabla}\times\vec{B})\cdot(\vec{E}-\eta\vec{J})
\end{equation}
After using mass conservation, we obtain:
\ber
  & &\frac{\partial}{\partial t}[\rho(\frac{1}{2}v^2+\phi+\epsilon)]+
     \vec{\nabla}\cdot[\rho\vec{v}(\frac{1}{2}v^2+\phi+\epsilon)+p\vec{v}-k\vec{\nabla}T]
     \nonumber \\
 &=&\rho T\frac{d s}{dt}+{v_i}\cdot\frac{\partial \sigma_{ik}^\prime}{\partial x_k}+
          \frac{1}{4\pi}(\vec{\nabla}\times\vec{B})\cdot \vec{E}
    -\frac{1}{4\pi}(\vec{\nabla}\times\vec{B})\cdot\eta\vec{J}-\vec{\nabla}\cdot(k\vec{\nabla}T)
    \label{energcon1}
\enr

\subsubsection{Magnetic Energy}

The temporal derivative of the magnetic field energy is derived using \ern{Fareq}:
\begin{equation}
    \frac{\partial}{\partial t}(\frac{B^2}{8\pi})=\frac{1}{4\pi}\vec{B}\cdot\frac{\partial\vec{B}}{\partial t}= - \frac{1}{4\pi}\vec{B}\cdot \vec{\nabla}\times\vec{E}
\label{magen2}
\end{equation}

\subsubsection{Total energy}

Now using the vector identity:
\beq
\vec{\nabla}\cdot(\vec{E}\times\vec{B})=
\vec{B}\cdot\vec{\nabla}\times\vec{E}-\vec{E}\cdot\vec{\nabla}\times\vec{B}
\label{veciden}
\enq
and \ern{magen2} in \ern{energcon1} we arrive at the following form:
\ber
& &    \frac{\partial}{\partial t}[\rho(\frac{1}{2}v^2+\phi+\epsilon)+\frac{B^2}{8\pi}]+
    \vec{\nabla}\cdot[\rho\vec{v}(\frac{1}{2}v^2+\phi+\epsilon+\frac{p}{\rho})
    -k\vec{\nabla}T+\frac{1}{4\pi}\vec{E}\times\vec{B}]
\nonumber \\
&=&\rho T\frac{d s}{dt}+{v_i}\cdot\frac{\partial \sigma_{ik}^\prime}{\partial x_k}
    -\frac{1}{4\pi}(\vec{\nabla}\times\vec{B})\cdot\eta\vec{J}-\vec{\nabla}\cdot(k\vec{\nabla}T)
\label{energcon2}
\enr
here $w = \epsilon+\frac{p}{\rho}$ is specific enthalpy and $\frac{1}{4\pi}\vec{E}\times\vec{B}$ is Poynting vector. We can write:
\beq
{v_{i}}\cdot\frac{\partial \sigma_{ik}^\prime}{\partial x_k}= \vec \nabla \cdot(\vec{v}\cdot\sigma^\prime)-\sigma_{ik}^\prime\frac{\partial v_{i}}{\partial x_{k}}
\label{viscten}
\enq
notice that as $\vec v$ is a vector and $\sigma^\prime$ is a tensor it follows
that $v_i \sigma_{ik}^\prime$ is  a vector. Therefore the total energy equation is:
\ber
  & &   \frac{\partial}{\partial t}[\rho(\frac{1}{2}v^2+\phi+\epsilon)+\frac{B^2}{8\pi}]+
    \vec{\nabla}\cdot[\rho\vec{v}(\frac{1}{2}v^2+
    \phi+w)-\vec{v}\cdot\sigma^\prime-k\vec{\nabla}T+\frac{1}{4\pi}\vec{E}\times\vec{B}]
    \nonumber \\
  &=&\rho T\frac{d s}{dt}-\sigma_{ik}^\prime\frac{\partial v_{i}}{\partial x_{k}}
    -\frac{1}{4\pi}(\vec{\nabla}\times\vec{B})\cdot\eta\vec{J}-\vec{\nabla}\cdot(k\vec{\nabla}T)
\label{energcon3}
    \enr
Energy conservation requires that the change of energy in any given infinitesimal volume is
equal to the net energy flux incoming and outgoing into the said volume. Hence, the two
terms in the left hand side of the equation must add up to zero, since one describes the temporal change of energy and the other all possible processes that contribute to the energy flux (convection, heat conduction and radiation). Thus \cite{LANDAU1987192,ogilvie2016astrophysical}, the right hand side of the equation should also be zero. Therefore:
\begin{equation}
    \rho T\frac{d s}{dt}=\sigma_{ik}^\prime\frac{\partial v_{i}}{\partial x_{k}}+\frac{1}{4\pi}(\vec{\nabla}\times\vec{B})\cdot\eta\vec{J}+\vec{\nabla}\cdot(k\vec{\nabla}T)
\label{entropyeq}
\end{equation}

Therefore:
\begin{equation}
    \rho T\frac{d s}{dt}=\sigma_{ik}^\prime\frac{\partial v_{i}}{\partial x_{k}}+\eta J^2+\vec{\nabla}\cdot(k\vec{\nabla}T)
\label{entropyeq2}
\end{equation}
This is the equation for rate of change of entropy in non-ideal MHD case which is not conserved. If there is no magnetic field in the system, it is exactly  the equation of heat transfer given by \cite{LANDAU1987192}. Non conservation of entropy depends on the all dissipative and non ideal processes present in the system. If we put $\mu$, $\eta$ and $k$ equal to zero, we will recover the ideal case i.e., $\frac{ds}{dt}=0$. Also notice \cite{LANDAU1987192} that the total
entropy of the system cannot decrease as the terms in the right hand side will lead only to an increase of the total entropy.

At this point we have set of basic equations (\ref{eqofmo},\ref{cont},\ref{magfl},\ref{Bdiff}) and (\ref{entropyeq2}) in non-ideal MHD and are ready to study the conservation of cross helicity.

\section{Cross Helicity for Non-Ideal Non-Barotropic MHD}

In this section, we derive the time derivative of non barotropic cross helicity using the aforementioned equations.

\subsection{A brief explanation of the topological velocity and non barotropic cross helicity}

The mathematical expression for cross helicity of non-barotropic fluids is given by \cite{webb2014localAA,yahalom2017conserved,yahalom2017non}:
\begin{equation}
    H_{CNB} =\int d^3x\, \vec{v}_t \cdot \vec{B}.
\label{HCNB}
\end{equation}
Here the topological velocity field is defined as $\vec{v_t}=\vec{v}-\sigma\vec{\nabla}s$ \cite{yahalom2021noether}, $\sigma$ is auxiliary variable, which depends on the Lagrangian time integral of the temperature i.e.:
\beq
\frac{d\sigma}{dt}=T.
\label{sigmadef}
\enq
Notice that in non-barotropic MHD one can calculate the temporal derivative of the cross helicity \ern{CH} using the standard equations and obtain:
\begin{equation}
    \frac{dH_C}{dt}=\int T\vec{\nabla}s\cdot\Vec{B}\hspace{0.1cm}d^3 x
\label{CH0}
\end{equation}
\\
So, generically cross helicity is not conserved.
A clue on how to define cross helicity for nonbarotropic MHD can be obtained from the variational analysis described in \cite{yahalom2016simplified} which is valid for magnetic field lines at the intersection of two comoving surfaces $\chi$, $\eta_0$ (Euler potentials). Following \cite{sakurai1979new} the magnetic field takes the form:
\begin{equation}
    \vec{B}=\vec{\nabla}\chi\times\vec{\nabla}\eta_0
    \label{MAG}
\end{equation}
And the generalized Clebsch representation of the velocity \cite{yahalom2016simplified} is:
\begin{equation}
    \vec{v}=\vec{\nabla}\nu+\alpha\vec{\nabla}\chi+\beta\vec{\nabla}\eta_0+\sigma\vec{\nabla}s
    \label{velocity}
\end{equation}
in the above $\alpha,\beta$ and $\nu$ are Lagrange multipliers appearing in the said action \cite{yahalom2016simplified}. Let us now write the cross helicity given in \ern{CH} in terms of \ern{MAG} and \ern{velocity}, this will take the form:
\begin{equation}
    H_C=\int d\Phi[\nu]+\int d\Phi\oint\sigma ds
\label{CH00}
\end{equation}
in which $d\Phi = \vec B \cdot d \vec S$ is the magnetic flux, and $[\nu]$ is the discontinuity of the non single valued potential $\nu$ \cite{yahalom2017conserved}. Now as for ideal MHD, the magnetic field lines move with the flow, it follows that the magnetic flux $d\Phi$ is conserved.
It is also shown in \cite{yahalom2017conserved} that the material derivative of $[\nu]$ must vanish.
Thus the first term in the right hand side of \ern{CH00} is conserved. This suggests the following definition for the non-barotropic cross helicity $H_{CNB}$:
\begin{equation}
    H_{CNB}=\int d\Phi[\nu]=H_C-\int d\Phi\oint\sigma ds
    \label{CNB00}
\end{equation}
The conventional form of the same expression is given in \ern{HCNB}. Please refer to \cite{yahalom2017conserved} for the detailed justification for the definition, and form of non-barotropic cross helicity and a proof of its constancy.

\subsection{The temporal derivative of non barotropic cross helicity}

Next we study the temporal derivative of the non-barotropic cross-helicity:
\begin{equation}
    \frac {d H_{CNB}}{dt} =\int d^3 x\,(\vec{v}_t \cdot \frac {\partial \vec{B}}{\partial t} +\vec{B} \cdot \frac {\partial \vec{v}_t}{\partial t}).
\label{dHCNB}
\end{equation}
Now we calculate first term on RHS with the help of \ern{Bdiff}:
\begin{equation}
    \vec{v}_t \cdot\frac {\partial \vec{B}}{\partial t}=\vec{v}_t\cdot \{\vec{\nabla} \times (\vec{v}\times\vec{B}) +\frac{\eta}{4\pi}\nabla^2\vec{B}\}
\label{dHCNB2}
\end{equation}
\begin{equation}
    \vec{v}_t \cdot\frac {\partial \vec{B}}{\partial t}=\vec{\nabla}\cdot \{(\vec{v}\times\ \vec{B})\times \vec{v}_t\}+(\vec{v}\times\ \vec{B})\cdot\vec{\omega}_t+\vec{v}_t\cdot\frac{\eta}{4\pi}\nabla^2\vec{B}\
\label{dHCNB3}
\end{equation}
where we define the topological vorticity of the topological flow field as:
\beq
\vec{\omega}_t \equiv \vec{\nabla}\times \vec{v}_t
\label{vortop}
\enq
Next we calculate second term on RHS:
\begin{equation}
    \ \vec{B}\cdot\frac {\partial \vec{v}_t}{\partial t}=\ \vec{B}\cdot\frac {\partial (\vec{v}-\sigma\vec{\nabla}\ s)}{\partial t}= \vec{B}\cdot(\frac{\partial \vec{v}}{\partial t}-\frac{\partial \sigma}{\partial t}\vec{\nabla} s-\sigma\vec{\nabla}\frac{\partial s}{\partial t})
\label{dHCNB4}
\end{equation}
Now we simplify right hand side of \ern{dHCNB4} in three steps: The first term is calculated with the help of \ern{eqofmo}:
\ber
    \frac{\partial \vec{v}}{\partial\ t}&=&-(\vec{v}\cdot\vec{\nabla})\vec{v}-\frac{\vec{\nabla} p}{\rho}+\frac{(\vec{\nabla}\times\ \vec{B})\times\ \vec{B}}{4\pi\rho}-\vec{\nabla}\phi+\frac{1}{\rho}\frac{\partial \sigma_{ik}^\prime}{\partial x_k}
\nonumber \\
    &=&(\vec{v}\times\vec{\omega})+\frac{(\vec{\nabla}\times\ \vec{B})\times\ \vec{B}}{4\pi\rho}-\vec{\nabla}(\frac{v^2}{2})-\vec{\nabla} w+T\vec{\nabla} s-\vec{\nabla}\phi+\frac{1}{\rho}\frac{\partial \sigma_{ik}^\prime}{\partial x_k}
\label{eqofmo2}
\enr
in which the vorticity is:
\beq
\vec{\omega} \equiv \vec{\nabla}\times \vec v
\label{vort}
\enq
and we have used the thermodynamical identity:
\beq
dw = d\varepsilon + d(\frac{p}{\rho}) = T ds +  \frac{1}{\rho} dp
\Rightarrow
\vec \nabla w =  T \vec \nabla s +  \frac{1}{\rho} \vec \nabla p.
\label{thermodyn}
\enq
Thus:
\begin{equation}
    \vec{B}\cdot\frac{\partial \vec{v}}{\partial\ t}=\vec{B}\cdot\left\{(\vec{v}\times\vec{\omega})-\vec{\nabla}(\frac{v^2}{2}+w+\phi)+T\vec{\nabla} s\right\}+\frac{B_i}{\rho}\frac{\partial \sigma_{ik}^\prime}{\partial x_k}.
\label{dHCNB5}
\end{equation}
In the second term we use \ern{sigmadef} to obtain:
\begin{equation}
 -\frac{\partial\sigma}{\partial t}\vec{\nabla} s=(\vec{v}\cdot\vec{\nabla}\sigma-T)\vec{\nabla} s.
 \label{dHCNB6}
\end{equation}
In the third term we use \ern{entropyeq2} to derive:
\begin{equation}
    -\sigma\vec{\nabla}\frac{\partial s}{\partial t}=\sigma\vec{\nabla}[\vec{v}\cdot\vec{\nabla}s-\frac{1}{\rho T}\sigma_{ik}^\prime\frac{\partial v_{i}}{\partial x_{k}}-\frac{\eta}{\rho T}J^2-\frac{1}{\rho T}\vec{\nabla}\cdot(k\vec{\nabla}T)]
\label{dHCNB7}
\end{equation}
Combining the above expressions, we have:
\ber
    \vec{B}\cdot\frac{\partial \vec{v}_t}{\partial\ t}&=&\vec{B}\cdot\left[(\vec{v}\times\vec{\omega})-
    \vec{\nabla}(\frac{v^2}{2}+w +\phi)+(\vec{v}\cdot\vec{\nabla}\sigma)\vec{\nabla} s+\sigma\vec{\nabla}(\vec{v}\cdot\vec{\nabla} s) \right.
\nonumber \\
    &+& \left. \sigma\vec{\nabla}\{-\frac{1}{\rho T}\sigma_{ik}^\prime\frac{\partial v_{i}}{\partial x_{k}}-\frac{\eta}{\rho T}J^2-\frac{1}{\rho T}\vec{\nabla}\cdot(k\vec{\nabla}T)\}\right]+\frac{B_i}{\rho}\frac{\partial \sigma_{ik}^\prime}{\partial x_k}.
\label{dHCNB8}
\enr
Notice that:
\beq
\vec \nabla \{\sigma(\vec{v}\cdot\vec{\nabla} s)\}  =
\sigma \vec \nabla (\vec v\cdot\vec \nabla s)+
(\vec v\cdot\vec \nabla s ) \vec \nabla  \sigma,
\label{dHCNB8b}
\enq
and also that:
\beq
\vec v \times \vec \omega_t = \vec v \times (\vec \omega - \vec \nabla  \sigma \times \vec \nabla s)
= \vec v \times \vec \omega + (\vec v \cdot \vec \nabla  \sigma ) \vec \nabla s
- (\vec v \cdot \vec \nabla  s ) \vec \nabla \sigma
= \vec v \times \vec \omega + (\vec v \cdot \vec \nabla  \sigma ) \vec \nabla s
- \vec \nabla ((\vec v \cdot \vec \nabla  s ) \sigma ) +\sigma \vec \nabla (\vec v \cdot \vec \nabla  s )
\label{dHCNB8c}
\enq
or that:
\beq
\vec v \times \vec \omega_t+\vec \nabla ((\vec v \cdot \vec \nabla  s ) \sigma )
= \vec v \times \vec \omega + (\vec v \cdot \vec \nabla  \sigma ) \vec \nabla s
+\sigma \vec \nabla (\vec v \cdot \vec \nabla  s )
\label{dHCNB8d}
\enq
Thus we obtain:
\beq
    \vec{B}\cdot\frac{\partial\vec{v}_t}{\partial\ t}=
    \vec{B}\cdot\left[(\vec{v}\times\vec{\omega}_t)+\vec{\nabla}\{\sigma(\vec{v}\cdot\vec{\nabla} s)-\frac{v^2}{2}-w-\phi\}
    -\sigma\vec{\nabla}\{\frac{1}{\rho T}\sigma_{ik}^\prime\frac{\partial v_{i}}{\partial x_{k}}+\frac{\eta}{\rho T}J^2+\frac{1}{\rho T}\vec{\nabla}\cdot(k\vec{\nabla}T)\}\right]+
    \frac{B_i}{\rho}\frac{\partial \sigma_{ik}^\prime}{\partial x_k}
\label{dHCNB9}
\enq
Combining \ern{dHCNB3} and \ern{dHCNB9} and taking into account that:
\beq
    \vec{B}\cdot (\vec{v}\times\vec{\omega}_t) =
     - (\vec{v}\times \vec{B}) \cdot \vec{\omega}_t
\label{dHCNB9b}
\enq
we obtain:
\ber
   & & \vec{v}_t\cdot\frac{\partial \vec{B}}{\partial t}+\frac{\partial \vec{v}_t}{\partial t}\cdot \vec{B}=\vec{\nabla}\cdot\{(\vec{v}\times\ \vec{B})\times \vec{v}_t\}+\frac{\eta}{4\pi}\vec v_t\cdot\nabla^2 \vec{B}+\vec{B}\cdot\vec{\nabla}\{\sigma(\vec{v}\cdot\vec{\nabla} s)-\frac{v^2}{2}-w-\phi\}
    \nonumber \\
    &+&\frac{B_i}{\rho}\frac{\partial \sigma_{ik}^\prime}{\partial x_k}- \sigma\vec{B}\cdot\vec{\nabla}\left\{\frac{1}{\rho T}\sigma_{ik}^\prime\frac{\partial v_{i}}{\partial x_{k}}+\frac{\eta}{\rho T}J^2+\frac{k}{\rho T}\nabla^2 T \right\}
\nonumber \\
    &=& \vec \nabla \cdot \left[\{(\vec{v}\times\ \vec{B})\times \vec{v}_t\} +\vec{B}\{\sigma(\vec{v}\cdot\vec{\nabla}s)-\frac{v^2}{2}-w-\phi \}
    - \sigma \vec B \{\frac{1}{\rho T}\sigma_{ik}^\prime\frac{\partial v_{i}}{\partial x_{k}}+\frac{\eta}{\rho T}J^2+\frac{k}{\rho T}\nabla^2 T \}
    \right]
\nonumber \\
    &+&\frac{\eta}{4\pi}\vec v_t\cdot\nabla^2 \vec{B}+\frac{B_i}{\rho}\frac{\partial \sigma_{ik}^\prime}{\partial x_k}
+ (\vec B \cdot \vec \nabla \sigma) \left[ \frac{1}{\rho T}\sigma_{ik}^\prime\frac{\partial v_{i}}{\partial x_{k}}+\frac{\eta}{\rho T}J^2+\frac{k}{\rho T}\nabla^2 T \right]
\label{dHCNB10}
\enr
Now substituting \ern{dHCNB10} into \ern{dHCNB}, we obtain:
\ber
 & &   \frac{d H_{CNB}}{dt}=
 \nonumber \\
 & &\int \vec \nabla \cdot \left[\{(\vec{v}\times\ \vec{B})\times \vec{v}_t\} +\vec{B}\{\sigma(\vec{v}\cdot\vec{\nabla}s)-\frac{v^2}{2}-w-\phi \}
    - \sigma \vec B \{\frac{1}{\rho T}\sigma_{ik}^\prime\frac{\partial v_{i}}{\partial x_{k}}+\frac{\eta}{\rho T}J^2+\frac{k}{\rho T}\nabla^2 T \}
    \right]d^3x +
 \nonumber \\
 & & \int \left\{ \frac{\eta}{4\pi}\vec v_t\cdot\nabla^2 \vec{B}+\frac{B_i}{\rho}\frac{\partial \sigma_{ik}^\prime}{\partial x_k}
+ (\vec B \cdot \vec \nabla \sigma) \left[ \frac{1}{\rho T}\sigma_{ik}^\prime\frac{\partial v_{i}}{\partial x_{k}}+\frac{\eta}{\rho T}J^2+\frac{k}{\rho T}\nabla^2 T \right]\right\}d^3x
\label{dHCNB11}
\enr
using Gauss divergence theorem, we obtain:
\ber
    & &   \frac{d H_{CNB}}{dt}=
 \nonumber \\
 & &\oint \left[\{(\vec{v}\times\ \vec{B})\times \vec{v}_t\} +\vec{B}\{\sigma(\vec{v}\cdot\vec{\nabla}s)-\frac{v^2}{2}-w-\phi \}
    - \sigma \vec B \{\frac{1}{\rho T}\sigma_{ik}^\prime\frac{\partial v_{i}}{\partial x_{k}}+\frac{\eta}{\rho T}J^2+\frac{k}{\rho T}\nabla^2 T \}
    \right]\cdot d \vec S +
 \nonumber \\
 & & \int \left\{ \frac{\eta}{4\pi}\vec v_t\cdot\nabla^2 \vec{B}+\frac{B_i}{\rho}\frac{\partial \sigma_{ik}^\prime}{\partial x_k}
+ (\vec B \cdot \vec \nabla \sigma) \left[ \frac{1}{\rho T}\sigma_{ik}^\prime\frac{\partial v_{i}}{\partial x_{k}}+\frac{\eta}{\rho T}J^2+\frac{k}{\rho T}\nabla^2 T \right]\right\}d^3x
\label{dHCNB12}
\enr
Here, the surface integral encapsulates the volume for which the cross helicity is calculated. If the surface is taken at infinity the magnetic fields vanish and thus in a generic case the entire surface term. Hence the time derivative of cross helicity can be written as:
\beq
    \frac{d H_{CNB}}{dt}=\int\left\{ \frac{\eta}{4\pi}\vec v_t\cdot\nabla^2 \vec{B}+\frac{B_i}{\rho}\frac{\partial \sigma_{ik}^\prime}{\partial x_k}
+ (\vec B \cdot \vec \nabla \sigma) \left[ \frac{1}{\rho T}\sigma_{ik}^\prime\frac{\partial v_{i}}{\partial x_{k}}+\frac{\eta}{\rho T}J^2+\frac{k}{\rho T}\nabla^2 T \right]\right\}d^3x
 \label{dHCNB13}
\enq
Thus, time derivative of cross helicity depends generically on the stress tensor i.e., viscosity of the fluid and coefficient of magnetic diffusivity, and also heat conduction but not on heat convection. By putting all non ideal terms to zero, we obtain the ideal MHD condition and conservation of non barotropic cross helicity takes place. In the special case that the magnetic field lies on $\sigma$ surfaces and thus is orthogonal to $\vec \nabla \sigma$,
the cross helicity change will not depend of thermal conductivity:
\beq
    \frac{d H_{CNB}}{dt}=\int\left\{ \frac{\eta}{4\pi}\vec v_t\cdot\nabla^2 \vec{B}+\frac{B_i}{\rho}\frac{\partial \sigma_{ik}^\prime}{\partial x_k}
\right\}d^3x
 \label{dHCNB14}
\enq
 The same will be true for
a high density and high temperature plasma, and a plasma of small temperature gradients (plasma in global thermal equilibrium). Of course even if heat conduction does not affect the non barotropic cross helicity, other non ideal processes do, those include friction and ohmic losses.

To conclude this subsection we shall partition the time derivative of the non barotropic cross helicity in accordance with the non ideal process that contributes to its modification:
\ber
    \frac{d H_{CNB}}{dt}&=&\int\left\{
     \eta \left[ \frac{\vec v_t \cdot \nabla^2 \vec B}{4\pi} + \frac{J^2}{\rho T} (\vec B \cdot \vec \nabla \sigma) \right] \right.
     \nonumber \\
     &+& k (\vec B \cdot \vec \nabla \sigma)\frac{\nabla^2 T}{\rho T}
    + \left. \frac{B_i}{\rho}\frac{\partial \sigma_{ik}^\prime}{\partial x_k}
+ (\vec B \cdot \vec \nabla \sigma)  \frac{\sigma_{ik}^\prime}{\rho T}\frac{\partial v_{i}}{\partial x_{k}}\right\}d^3x
 \label{dHCNB15}
\enr
Let us introduce the dimensionless Reynolds number and magnetic Reynolds number:
\beq
R_e \equiv \frac{\bar \rho U L}{\mu}, \qquad R_m \equiv \frac{U L}{\eta}
 \label{Reynolds}
\enq
where $L$ is a characteristic length, $\bar \rho$ is a typical density and $U$ a characteristic speed of the system.

We may now inquire how does the value of those  numbers affect the conservation of non barotropic cross helicity. To do this we write each physical variable $g$ as a multiplication of a characteristic value $\bar g$ and a dimensionless variable $g'$ in the form:
\beq
g =  \bar g g' \quad \Rightarrow \quad \vec x = L  \vec x', \quad \vec v = U  \vec v', \quad t = \bar t t'
\label{dimensionless}
\enq
the above equation suggest the following choice of $\bar t$:
\beq
\bar t \equiv \frac{L}{U}.
\label{bart}
\enq
Similarly we write:
\beq
\rho = \bar \rho \rho', \quad  T = \bar T T', \quad  \sigma = \bar \sigma \sigma', \quad \vec B = \bar B \vec B', \quad \vec J = \bar J \vec J'
\label{dimensionless2}
\enq
\Er{sigmadef} suggests the following definition of $\bar \sigma$:
\beq
\bar \sigma \equiv \bar T~\bar t = \frac{\bar T L}{U}.
\label{bsigma}
\enq
and \ern{ampere} suggests the following definition of $\bar J$:
\beq
\bar J \equiv \frac{\bar B}{L}.
\label{bJ}
\enq
For the viscosity tensor we use a double prime notation (as we already used a single prime notation
to distinguish the viscosity tensor from the $\sigma$ scalar). Thus:
\beq
\sigma_{ik}^{'} = \bar \sigma' \sigma_{ik}^{''}, \qquad
\sigma_{ik}^{''}\equiv \frac{\partial v'_i}{\partial x'_k}+\frac{\partial v'_k}{\partial x'_i}-\frac{2}{3}\delta_{ik}\frac{\partial v'_l}{\partial x'_l}.
\label{bsigma1}
\enq
It follows from \ern{visten} that:
\beq
\bar \sigma' =  \frac{\mu U}{L} = \frac{\bar \rho U^2}{R_e}.
\label{bsigma2}
\enq
We will also need the magnetic and kinetic energy expressions in dimensionless form:
\ber
E_k &=& \frac{1}{2} \int \rho \vec v^2 d^3 x = \bar \rho U^2 L^3 E'_k, \qquad
E'_k \equiv \frac{1}{2} \int \rho' \vec v'^2 d^3 x'
\nonumber \\
E_m &=& \frac{1}{8 \pi} \int \vec B^2 d^3 x = \bar B^2 L^3 E'_m, \qquad
E'_m \equiv \frac{1}{8 \pi} \int \vec B'^2 d^3 x'.
\label{Kinmagenerg}
\enr
Finally we shall look at the amount of heat $Q_c$ that is conducted into the volume (which is not
equal to the total change in heat in the volume as heat may be produced by viscosity and ohmic losses see \ern{entropyeq}). We may write the heat flux density given in \ern{qheatflux} as:
\beq
 \vec q = \bar q \vec q',  \qquad   \vec q' \equiv \vec{\nabla'} T', \qquad \bar q \equiv k \frac{\bar T}{L}.
 \label{dimlessq}
\enq
Thus the rate of change of $Q_c$ is (in dimensional and dimensionless form):
\beq
 \frac{d Q_c}{d t} = -\oint \vec q \cdot d \vec S = - L^2 \bar q \oint \vec q' \cdot d \vec S', \qquad  \frac{d Q'_c}{d t'} = -\oint \vec q' \cdot d \vec S'
 \label{Qcch}
\enq
where we integrate over a surface encapsulating the flow. Writing as usual:
\beq
 \bar Q_c = \frac{Q_c}{Q'_c} = \bar q \frac{L^3}{U} = k \bar T \frac{L^2}{U}
 \label{Qcch2}
\enq
Having defined the above quantities we may write the non barotropic cross helicity as:
\beq
H_{CNB} = \bar H_{CNB} H'_{CNB}, \qquad  \bar H_{CNB} \equiv  L^3 U \bar B, \qquad
 H'_{CNB} \equiv \int d^3 x'\, \vec{v'}_t \cdot \vec{B'}.
\label{HCNBdimless}
\enq
Thus:
\beq
\frac{d H_{CNB}}{dt} = L^2 U^2 \bar B \frac{d H'_{CNB}}{dt'} \quad \Rightarrow \quad
\frac{d H'_{CNB}}{dt'} = \frac{1}{L^2 U^2 \bar B} \frac{d H_{CNB}}{dt}.
\label{dHCNBdimless}
\enq
It follows that \ern{dHCNB15} can be written in the form:
\ber
 \frac{d H'_{CNB}}{dt'} &=& \frac{1}{L^2 U^2 \bar B} \frac{d H_{CNB}}{dt} = \int\left\{
     \frac{1}{R_m} \left[ \frac{\vec v'_t \cdot \nabla'^2 \vec B'}{4\pi} + \frac{J'^2}{\rho' T'} \vec B' \cdot \vec \nabla' \sigma' \frac{E'_k}{E'_m} \frac{E_m}{E_k} \right] \right.
     \nonumber \\
 &+& \frac{E'_k}{Q'_c} \frac{Q_c}{E_k} (\vec B' \cdot \vec \nabla' \sigma')\frac{\nabla'^2 T'}{\rho' T'}
    + \left. \frac{1}{R_e} \left[\frac{B'_i}{\rho'}\frac{\partial \sigma_{ik}^{''}}{\partial x'_k}
+ (\vec B' \cdot \vec \nabla' \sigma')  \frac{\sigma_{ik}^{''}}{\rho' T'}\frac{\partial v'_{i}}{\partial x'_{k}}\right] \right\}d^3x'
 \label{dHCNB15b}
\enr
thus generally speaking, non barotropic cross helicity will change slowly for flows
with both high Reynolds and high magnetic Reynolds numbers in which the heat conducted is
small with respect to the kinetic energy of the flow. Indeed "Increasing cross helicity with fixed fluctuation energy increases the time required for energy to cascade to smaller scales, reduces the cascade power, and increases the anisotropy of the small-scale fluctuations" \cite{Chandran}. This has implications for the solar wind and solar corona \cite{Chandran}.

Similarly if we have a look on magnetic helicity in brief, it is defined as:
\beq
H_M \equiv \int \vec{A} \cdot \vec{B} ~ d^3 x,
\label{HM}
\enq
where $\vec A$ is the magnetic vector potential, defined such that:
\beq
\vec B=\vec \nabla \times \vec A.
\label{AB}
\enq
For an illustration of specific helical magnetic fields see \cite{yahalom2008simplified}.
After taking the temporal derivative of the magnetic helicity and simplifying the expressions, we obtain the well known relation:
\begin{equation}
\frac{dH_M}{dt}=-2\eta\int \vec{J}\cdot\vec{B}\ d^3 x.
\label{dHM23}
\end{equation}
The above relation has been verified by many authors \cite{biskamp1997nonlinear,priest2014magnetohydrodynamics,akhmet2009remark,verma2019energy,Verma_2021}. It is clear from \ern{dHM23} that generally speaking the magnetic diffusivity leads to the non conservation of magnetic helicity in non-ideal MHD. On the other hand nor viscosity nor heat conductivity affect the conservation of magnetic helicity. We can easily recover the ideal MHD condition by putting magnetic diffusivity to zero, in which case it is evident that the magnetic helicity is conserved.
Furthermore, the above result also shows that magnetic helicity is conserved even in non ideal flows if the currents are orthogonal to the magnetic field i.e., $\vec{J}\cdot\vec{B}=0$.
In this case:
\begin{equation}
\frac{dH_M}{dt}=0.
\label{dHM24}
\end{equation}
If the magnetic field and magnetic current density are not strictly orthogonal then the magnetic helicity is only approximately conserved.

The above result can also be expressed in a dimensionless form in which we write the vector potential as:
\beq
\vec A = \bar A \vec A', \qquad \bar A = \bar B L.
\label{ABdimle}
\enq
And thus:
\beq
H_M  = \bar H_M H'_M, \qquad  H'_M \equiv \int \vec{A'} \cdot \vec{B'} ~ d^3 x',
\qquad  \bar H_M \equiv \bar B^2 L^4.
\label{HMdimle}
\enq
It is straightforward to see that:
\begin{equation}
\frac{dH'_M}{dt'}=-\frac{2}{R_m} \int \vec{J'}\cdot\vec{B'}\ d^3 x'.
\label{dHM23dimle}
\end{equation}
It follows that an approximate conservation of magnetic helicity is achievable at high magnetic Reynolds number and is independent on the values of the Reynolds number and total conducted heat.
Thus the effect of magnetic helicity conservation is much more general.

\section{An Application}

We shall deal with the application in two stages, in the first we describe an ideal MHD flow following \cite{yahalom2021noether} then we assume a small magnetic diffusivity $\eta$ (high magnetic Reynolds number) and discuss the implications on the flow. Finally we calculate the
magnetic and cross helicities and discuss their relative rate of change.

\subsection{The ideal case}

We introduce a set of  standard cylindrical coordinates $R,\phi,z$, $\hat{R},\hat{\phi},\hat{z}$ are the corresponding unit vectors. We further assume an  MHD flow of uniform density $\rho$
confined between an internal $a_{in}$ and external $a$ radii: $a_{in} < R < a$. Furthermore assume that the flow contains a helical stratified stationary magnetic field:
\beq
\vec B = \left\{%
\begin{array}{ll}
   2 B_{\bot}(1-\frac{R}{a})\hat{\phi} + B_{z0}\hat{z} &  a_{in}<R<a \\
    0 & {\rm otherwise} \\
\end{array}%
\right.
\label{Bstra}
\enq
in which $B_{z0},B_{\bot}$ are constants. The magnetic field is contained
in a cylinder of Radius $a$ and is independent of $z$. Furthermore, we assume that the planes $z=0$ and $z=L$ can be identified such that a topological torus is created. In such a scenario the only field lines that will be closed will satisfy
the relation:
\beq
\frac{n}{m}=\frac{B_{\bot}}{\pi R B_{z0}} \left(1-\frac{R}{a}\right)L, \qquad n,m \ {\rm integers}
\label{closlin}
\enq
while lines not satisfying this relation will be surface filling.

In \cite{yahalom2021noether} we derive a stationary velocity field $\vec v$ that satisfy the stationary ideal versions of \er{Bdiff} and \er{cont}:
\beq
 \vec \nabla \times (\vec v \times \vec B) = 0
\label{Beqs}
\enq
\beq
\vec \nabla \cdot (\rho \vec v ) = 0
\label{masscons}
\enq
There we arrived at the simple expression:
\beq
\vec v = v_0 \frac{R}{a} \hat \phi,
\label{orthovstra}
\enq
$v_0$ is a constant with dimensions of velocity.
The ideal stationary version of \ern{eqofmo}
is given by:
\beq
\rho (\vec v \cdot \vec \nabla)\vec v  = -\vec \nabla p  + \frac{(\vec \nabla \times \vec B) \times \vec B}{4 \pi}
\label{Eulers}
\enq
This can be solved by the pressure function:
\beq
p(R) =  \frac{B_{\bot}^2}{\pi} \left(3 \frac{R}{a} - \frac{R^2}{a^2}-\ln (\frac{R}{a})-2\right) + \frac{1}{2}\rho  v_0^2 \left(\frac{R^2}{a^2}-1\right), \qquad p(a)=0.
\label{pressure}
\enq
of course $p(a_{in}) \neq 0$ and thus one will need a rigid internal cylinder of radius $a_{in}$ that can support such a pressure. We now can calculate the cross Helicity using \ern{HCNB} in which we assume uniform specific entropy such that $\vec v_t = \vec v$.
Inserting \ern{orthovstra} and \ern{Bstra} into \ern{HCNB} we arrive at the expression:
\beq
H_{CNBI}  =  \frac{\pi}{3} v_0 B_{\bot} \frac{L}{a^2} \left[a^4 - a_{in}^3(4a -3 a_{in})\right].
\label{Noether4c}
\enq
In order to calculate the magnetic helicity, one need to calculate a vector potential, one possibility is given by:
\beq
\vec A = \left\{%
\begin{array}{ll}
   \frac{1}{2} B_{z0} R \hat{\phi} -2 B_{\bot} R (1-\frac{R}{2 a}) \hat{z} & a_{in}<R<a \\
    0 & {\rm otherwise} \\
\end{array}%
\right.
\label{Astra}
\enq
Inserting \ern{Bstra} and \ern{Astra} into \ern{HM} we arrive at the result:
\beq
H_{MI} = - \frac{2}{3} \pi L B_{z0} B_{\bot} (a-a_{in}) \left[a^2 + a_{in} a +a_{in}^2 \right].
\label{MHstra}
\enq
Obviously both magnetic helicity and magnetic cross helicity do not change in time. The situation, however, is quite different when one considers non-ideal processes such as magnetic diffusion.

\subsection{The non-ideal case}

Let us assume a non-ideal magnetic diffusion, the magnetic field $\vec B_T$ will obviously be different from $\vec B$ and we may write it in the following form:
\beq
\vec B_T = \vec B + \eta \vec B_1
\label{BT}
\enq
In the above $\vec B$ is given in \ern{Bstra} and is thus a stationary solution of an ideal MHD configuration. If we take the typical scale to be $a$ and the typical velocity to be $v_0$
we can write the magnetic Reynolds number in the form:
\beq
R_m = \frac{v_0 a}{\eta}, \qquad \Rightarrow \qquad \eta =  \frac{v_0 a}{R_m}
\label{magRe}
\enq
Thus if we take the typical size of the magnetic fields $\vec B_T$ and $\vec B$ to be
$\bar B_T = \bar B =  B_{z0}$ and the typical size of $\vec B_1$ to be $\bar B_1 = \frac{B_{z0}}{v_0 a}$ we may write:
\beq
\vec B'_T = \vec B' + \frac{1}{R_m} \vec B'_1
\label{BTp}
\enq
We shall assume that the magnetic Reynolds number is large, such that the non-ideal correction is small. Now  $\vec B_T$ must satisfy \ern{Bdiff}:
\beq
\frac {\partial \vec{B_T}}{\partial t}=\vec{\nabla} \times (\vec{v}\times\ \vec{B_T}) +\frac{\eta}{4\pi}\nabla^2\vec{B_T}
\label{Bdiff2}
\enq
assuming that the velocity field is given by \ern{orthovstra}, and taking int account that $\vec B$ is stationary we arrive at:
\beq
\eta \frac {\partial \vec{B_1}}{\partial t}=\eta \vec{\nabla} \times (\vec{v}\times\ \vec{B_1}) +\frac{\eta}{4\pi}\nabla^2 (\vec{B} + \eta  \vec{B_1})
\label{B1dif}
\enq
Thus $\eta$ can be cancelled out and we obtain:
\beq
\frac {\partial \vec{B_1}}{\partial t}= \vec{\nabla} \times (\vec{v}\times\ \vec{B_1}) +\frac{1}{4\pi}\nabla^2 (\vec{B} + \eta  \vec{B_1})
\label{B1dif2}
\enq
the term $\eta  \vec{B_1} = \frac{B_{z0}}{R_m} \vec B'_1$ hence for high magnetic Reynolds numbers it can be neglected. Thus we arrive at the equation:
\beq
\frac {\partial \vec{B_1}}{\partial t}= \vec{\nabla} \times (\vec{v}\times\ \vec{B_1}) +\frac{1}{4\pi}\nabla^2 \vec{B}
\label{B1dif3}
\enq
the source terms of the above equation is according to \ern{Bstra}:
\beq
\frac{1}{4\pi}\nabla^2 \vec{B} = -\frac{B_{\bot}}{2 \pi R^2} \hat \phi
\label{B1difs}
\enq
for every point $R<a$. Moreover, it is easy to show that if at a specified time $t=0$  we have $\vec{B_1} (\vec x, 0) = 0$
it follows that $B_{1R} (\vec x, t) = B_{1z}  (\vec x, t) = 0$ for any time $t$. The equation for
$B_{1\phi}$ is thus:
\beq
\frac {\partial B_{1\phi}}{\partial t}= -\frac{B_{\bot}}{2 \pi R^2}
\label{B1dif4}
\enq
which can be trivially integrated:
\beq
B_{1\phi} = -\frac{B_{\bot}}{2 \pi R^2} t
\label{B1dif5}
\enq
Thus the total magnetic field is:
\beq
\vec B_T = \left\{%
\begin{array}{ll}
   2 B_{\bot}(1-\frac{R}{a}-\frac{\eta}{4 \pi R^2} t)\hat{\phi} + B_{z0}\hat{z} & a_{in}<R<a \\
    0 & {\rm otherwise} \\
\end{array}%
\right.
\label{BTstra}
\enq
And thus current density can be calculated using \ern{ampere} to be:
\beq
\vec J_T = \left\{%
\begin{array}{ll}
 \frac{B_{\bot}}{2 \pi R}  \left( 1-\frac{2R}{a}+ \frac{\eta}{4 \pi R^2} t \right) \hat z  & a_{in}<R<a \\
    0 & {\rm otherwise}  \\
\end{array}%
\right.
\label{JTstra}
\enq
Thus one can calculate the time dependent pressure using \ern{eqofmo}:
\beq
p(R,t) =  \frac{B_{\bot}^2}{\pi} \left(3 \frac{R}{a} - \frac{R^2}{a^2}-\ln (\frac{R}{a})-2
+ \frac{\eta t}{4 \pi a} (\frac{1}{R} - \frac{1}{a}) - \frac{\eta^2 t^2}{96 \pi^2} (\frac{1}{R^6} - \frac{1}{a^6})\right) +
 \frac{1}{2}\rho  v_0^2 \left(\frac{R^2}{a^2}-1\right), \qquad p(a,t)=0.
\label{pressuredif}
\enq
Thus the internal cylinder must sustain the above time dependent pressure.

\begin{figure}
\centering
\includegraphics[scale=0.85]{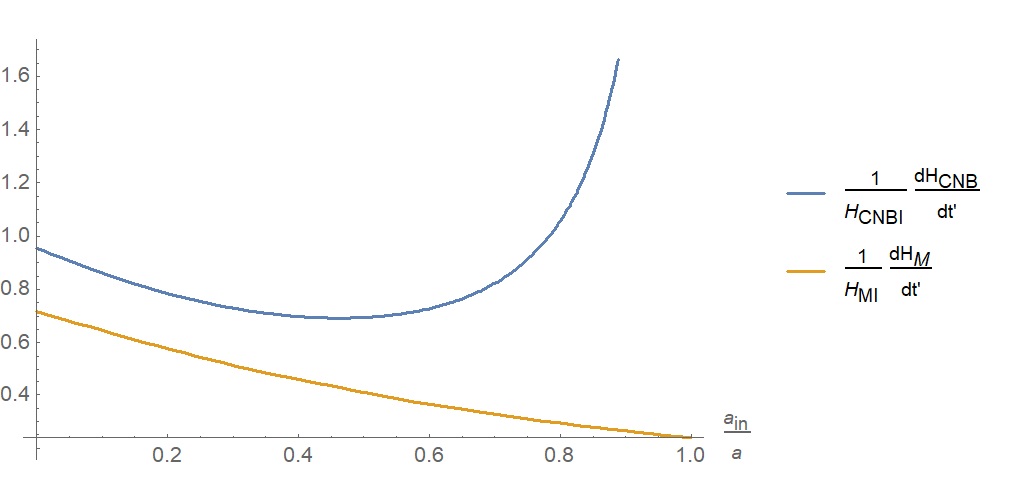}
\caption {Comparison $\left|\frac{1}{H_{MI}} \frac{d H_{MT}}{dt'}\right|$ and
$\left|\frac{1}{H_{CNBI}} \frac{d H_{CNBT}}{dt'}\right|$ for $R_m=1$}
\label{compofhelicons}
\end{figure}

Now to calculate the magnetic helicity we need a vector potential, this can be similarly obtained as in the ideal case in the form:
\beq
\vec A_T = \left\{%
\begin{array}{ll}
   \frac{1}{2} B_{z0} R \hat{\phi} -2 B_{\bot} R (1-\frac{R}{2 a}+\frac{\eta t}{4 \pi R^2}) \hat{z} & a_{in}<R<a \\
    0 & {\rm otherwise} \\
\end{array}%
\right.
\label{AstraT}
\enq
Inserting \ern{BTstra} and \ern{AstraT} into \ern{HM} will result in a time dependent magnetic helicity:
\beq
H_{MT} = - \frac{2}{3} \pi L B_{z0} B_{\bot} (a-a_{in}) \left[a^2 + a_{in} a +a_{in}^2 +\frac{9 \eta t}{4 \pi} \right].
\label{MHTstra}
\enq
The time derivative of the above magnetic helicity is:
\beq
\frac{d H_{MT}}{dt} = - \frac{3}{2} L B_{z0} B_{\bot} (a-a_{in}) \eta.
\label{dtMHTstra}
\enq
Similarly, by inserting \ern{BTstra} and \ern{orthovstra} into \ern{HCNB} will result in a time dependent cross helicity:
\beq
H_{CNBT}  =  4 \pi v_0 B_{\bot} \frac{L}{a} \left[\frac{1}{3}(a^3 -a_{in}^3)- \frac{1}{4a}
(a^4 -a_{in}^4) - \frac{\eta t}{4 \pi} (a -a_{in})\right].
\label{CNBT}
\enq
with a time derivative of:
\beq
\frac{d H_{CNBT}}{dt} = - v_0 B_{\bot} \frac{L}{a}  (a-a_{in}) \eta.
\label{dtCNBTstra}
\enq
It is interesting to compare which of the topological quantities is better preserved given
some high Reynolds number (and neglecting viscosity and heat conduction) to this end we compare:
\beq
\left|\frac{1}{H_{MI}} \frac{d H_{MT}}{dt'}\right|=\frac{9}{4 \pi R_m (1+a'_{in}+{a'}_{in}^2)},
\qquad  a'_{in} \equiv \frac{a_{in}}{a}, t' = t \frac{v_0}{a}
\label{dtMTfrac}
\enq
with:
\beq
\left|\frac{1}{H_{CNBI}} \frac{d H_{CNBT}}{dt'}\right|=
\frac{3 (1-a'_{in})}{\pi R_m \left[1 - {a'}_{in}^3(4-3{a'}_{in})\right]}.
\label{dtCNBTfrac}
\enq
Both the above quantities are inversely proportional to the magnetic Reynolds number $R_m$. The result for $Rm=1$ are depicted in figure \ref{compofhelicons}
which illustrates that the relative rate of change of both quantities is about the same
with a slight advantage to magnetic helicity depending on the particular geometry of the configuration.

\section{Summary and Concluding Remarks}

The conservation of topological constants of motion such as the non-barotropic cross helicity and magnetic helicity imposes constraints on the MHD flow and thus affect its stability \cite{yahalom2017non}. In his important review paper "Physics of magnetically confined plasmas" A. H. Boozer \cite{boozer2005physics} states that: "A spiky
current profile causes a rapid dissipation of energy relative to magnetic helicity. If the evolution of a magnetic field is rapid, then it must be at constant helicity." Recently
\cite{faraco2020proof} have given a proof of Taylor’s conjecture on magnetic helicity conservation, This, and other work, shows that magnetic helicity is well conserved, in turbulent flows, despite the presence of weak diffusion.

The application of the "Magnetic Aharonov-Bohm effect" is expected to be important in understanding the dynamics of magnetically confined plasmas and the problem of controlled fusion. Usually topological conservation laws are used in order to deduce lower bounds on the "energy" of the flow.
 Those bounds are only approximate in non ideal flows but due to their topological nature simulations show that  they are approximately conserved even when the "energy" is not.

 For example, \cite{moffatt1992relaxation} shows that the "energy"  is bounded from below by the magnetic helicity using the Cauchy-Schwarz inequality:
\begin{equation}
H_{M} = \int  \vec{B}\cdot \vec{A}d^{3} x \leq \sqrt{\int  \vec{A}^2 d^{3} x}\sqrt{\int \vec{B}^2 d^{3} x},
 \label{SchwarzHm}
\end{equation}
In addition it what shown by \cite{yahalom2017conserved}
 \begin{equation}
H_{M} = \int  \vec{B}\cdot \vec{A}d^{3} x \leq \frac{1}{2}\int  \left( \vec{B}^2 + \vec{A}^2 \right) d^{3} x,
 \label{Globhelboundm}
\end{equation}

 A similar analysis can be done for non-barotropic cross helicity. It is easy to show that the "energy"  is bounded from below by the cross helicity as follows:
 \begin{equation}
H_{CNB} = \int  \vec{B}\cdot \vec{v_t}d^{3} x \leq \frac{1}{2}\int  \left( \vec{B}^2 + \vec{v_t}^2 \right) d^{3} x,
 \label{Globhelbound}
\end{equation}
\begin{equation}
H_{CNB} = \int  \vec{B}\cdot \vec{v_t}d^{3} x \leq \sqrt{\int  \vec{v_t}^2 d^{3} x}\sqrt{\int \vec{B}^2 d^{3} x},
 \label{Globhelbound2}
\end{equation}
In this sense a configuration with a highly complicated topology is more stable since its energy is bounded from below.

The current paper highlights on the constancy of topological invariants when MHD flow is not ideal. To summarize, we have analytically examined the constancy of cross helicity for non-ideal compressible MHD by taking resistive heating, heat conduction and viscous effects into account. For achieving aforementioned aim, first we have derived a specific entropy equation which is valid in non-ideal MHD flow with the help of energy equation. After that, using the basic MHD equations and the definition of non barotropic cross helicity, we have derived the rate of change of generalized cross helicity with time. We have showed that the helicities are not conserved in non-ideal MHD flow and their time derivatives depends only on non ideal processes. One can easily recover ideal MHD conditions by putting all the non ideal constants to zero.

Magnetic and non barotropic cross helicities differ, however, in that magnetic helicity is only affected by magnetic diffusivity while non barotropic cross helicities is dependent on viscosity and heat conduction as well (except for special cases) making it more susceptible to change over time.
A dimensional analysis shows that for magnetic helicity "slow change" one needs only a high magnetic Reynolds number but this will not suffice for a "slow change" of non barotropic cross helicity. The later case requires in addition a high Reynolds number (non magnetic) as well as limitations to the amount of heat conducted with respect to the flows kinetic energy.

Applications of both topological invariants are widely known in solar plasma. Cross helicity and magnetic helicity are used to characterize the turbulence phenomena specially in astrophysics. Non-zero cross helicity has an important consequences related to transport and \cite{heinonen2021role} has showed the close relation between momentum transfer and cross helicity. \cite{perez2009role} has discussed its role in MHD turbulence by high resolution direct numerical simulations. Cross helicity is proportional to the correlation between velocity and magnetic field fluctuation and measures the relative importance of Alfv\'{e}n waves in global fluctuation and \cite{goldreich1995toward} have predicted the generation of interstellar turbulence caused by nonlinear interactions among Shear Alfv\'{e}n waves. Many studies in the literature have shown that  cross helicity is correlated to the self production of turbulence. Non-zero magnetic helicity may help to explain the presence of large scale magnetic structures in the universe. It is believed that it may induce the formation of magnetic structures at larger scales from smaller scale ones and this process can be referred as inverse cascading.
\cite{biskamp2003magnetohydrodynamic} has presented decay law for MHD turbulence with the help of conservation law for magnetic helicity. Variation in energy flux is also adopted by \cite{verma2004statistical} to study decay law in MHD turbulence.
\cite{mizeva2009cross} has discussed the effect of cross helicity on the cascade process in MHD turbulence. They showed from numerical results that "cross helicity blocks the spectral energy transfer in MHD turbulence and results in energy accumulation in the system. This accumulation proceeds until the vortex intensification compensates the decreasing efficiency of nonlinear interactions". The importance of cross helicity is also discussed by many other authors. \cite{smith2009turbulent} have adopted third-moment techniques
to study the scale of energy cascading in the solar wind at 1 AU with the assumption of isotropic turbulence. They find that "when the correlation between magnetic and velocity fluctuations is large there is a back transfer of energy within the spatial scales normally attributed to the inertial range where energy is moved from small to large scales in such a way as to reinforce the cross-field correlation. This may help to explain solar wind observations with large correlations indicative of outward-propagating fluctuations". Impact of non zero cross helicity on MHD turbulence is unparalleled and affects the global dynamics. \cite{briard2018decay} examined the effect of cross helicity on the decay of isotropic MHD turbulence and concluded that an initial non zero cross helicity makes imbalanced MHD turbulence. The subtle anisotropic effect of cross helicity can be the caused of this.

For better insight of possible applications in the case of non conservation for topological invariants, authors plan to develop variational principles for non ideal MHD in the future. Our aim is to use them for the analysis of the stability of non-ideal MHD configurations. As for designing efficient numerical schemes for integrating the equations of fluid dynamics and MHD, one may follow the approach described in \cite{yahalom2003method}. Analyzing the dynamics of the new generalized non-barotropic $\chi$ and $\eta$ cross helicities recently developed by \cite{yahalom2021noether} in the non ideal case will also be the part of future work as well as attention to the local forms of
magnetic and cross helicities.
\\ \\
{\bf Declaration of Interests: The authors report no conflict of interest.}
\\
\\

\bibliographystyle{apacite}
\bibliography{Ref}

\begin{thebibliography}{}

\bibitem [\protect \citeauthoryear {%
Akhmet’ev%
, Kunakovskaya%
\BCBL {}\ \BBA {} Kutvitskii%
}{%
Akhmet’ev%
\ \protect \BOthers {.}}{%
{\protect \APACyear {2009}}%
}]{%
akhmet2009remark}
\APACinsertmetastar {%
akhmet2009remark}%
\begin{APACrefauthors}%
Akhmet’ev, P\BPBI M.%
, Kunakovskaya, O\BPBI V.%
\BCBL {}\ \BBA {} Kutvitskii, V.%
\end{APACrefauthors}%
\unskip\
\newblock
\APACrefYearMonthDay{2009}{}{}.
\newblock
{\BBOQ}\APACrefatitle {Remark on the dissipation of the magnetic helicity
  integral} {Remark on the dissipation of the magnetic helicity
  integral}.{\BBCQ}
\newblock
\APACjournalVolNumPages{Theoretical and Mathematical
  Physics}{158}{1}{125--134}.
\PrintBackRefs{\CurrentBib}

\bibitem [\protect \citeauthoryear {%
Barnes%
\ \protect \BOthers {.}}{%
Barnes%
\ \protect \BOthers {.}}{%
{\protect \APACyear {1986}}%
}]{%
barnes1986experimental}
\APACinsertmetastar {%
barnes1986experimental}%
\begin{APACrefauthors}%
Barnes, C\BPBI W.%
, Fernandez, J.%
, Henins, I.%
, Hoida, H.%
, Jarboe, T.%
, Knox, S.%
\BDBL {}McKenna, K.%
\end{APACrefauthors}%
\unskip\
\newblock
\APACrefYearMonthDay{1986}{}{}.
\newblock
{\BBOQ}\APACrefatitle {Experimental determination of the conservation of
  magnetic helicity from the balance between source and spheromak}
  {Experimental determination of the conservation of magnetic helicity from the
  balance between source and spheromak}.{\BBCQ}
\newblock
\APACjournalVolNumPages{The Physics of fluids}{29}{10}{3415--3432}.
\PrintBackRefs{\CurrentBib}

\bibitem [\protect \citeauthoryear {%
Batchelor%
}{%
Batchelor%
}{%
{\protect \APACyear {1967}}%
}]{%
batchelor1967introduction}
\APACinsertmetastar {%
batchelor1967introduction}%
\begin{APACrefauthors}%
Batchelor, G\BPBI K.%
\end{APACrefauthors}%
\unskip\
\newblock
\APACrefYearMonthDay{1967}{}{}.
\newblock
{\BBOQ}\APACrefatitle {An Introduction to Fluid Dynamics,(1967)} {An
  introduction to fluid dynamics,(1967)}.{\BBCQ}
\newblock
\APACjournalVolNumPages{Cambridge,: UP xviii}{615}{}{}.
\PrintBackRefs{\CurrentBib}

\bibitem [\protect \citeauthoryear {%
Berger%
\ \BBA {} Field%
}{%
Berger%
\ \BBA {} Field%
}{%
{\protect \APACyear {1984}}%
}]{%
berger1984topological}
\APACinsertmetastar {%
berger1984topological}%
\begin{APACrefauthors}%
Berger, M\BPBI A.%
\BCBT {}\ \BBA {} Field, G\BPBI B.%
\end{APACrefauthors}%
\unskip\
\newblock
\APACrefYearMonthDay{1984}{}{}.
\newblock
{\BBOQ}\APACrefatitle {The topological properties of magnetic helicity} {The
  topological properties of magnetic helicity}.{\BBCQ}
\newblock
\APACjournalVolNumPages{Journal of Fluid Mechanics}{147}{}{133--148}.
\PrintBackRefs{\CurrentBib}

\bibitem [\protect \citeauthoryear {%
Biskamp%
}{%
Biskamp%
}{%
{\protect \APACyear {1997}}%
}]{%
biskamp1997nonlinear}
\APACinsertmetastar {%
biskamp1997nonlinear}%
\begin{APACrefauthors}%
Biskamp, D.%
\end{APACrefauthors}%
\unskip\
\newblock
\APACrefYear{1997}.
\newblock
\APACrefbtitle {Nonlinear magnetohydrodynamics} {Nonlinear
  magnetohydrodynamics}\ (\BNUM~1).
\newblock
\APACaddressPublisher{}{Cambridge University Press}.
\PrintBackRefs{\CurrentBib}

\bibitem [\protect \citeauthoryear {%
Biskamp%
}{%
Biskamp%
}{%
{\protect \APACyear {2003}}%
}]{%
biskamp2003magnetohydrodynamic}
\APACinsertmetastar {%
biskamp2003magnetohydrodynamic}%
\begin{APACrefauthors}%
Biskamp, D.%
\end{APACrefauthors}%
\unskip\
\newblock
\APACrefYear{2003}.
\newblock
\APACrefbtitle {Magnetohydrodynamic turbulence} {Magnetohydrodynamic
  turbulence}.
\newblock
\APACaddressPublisher{}{Cambridge University Press}.
\PrintBackRefs{\CurrentBib}

\bibitem [\protect \citeauthoryear {%
Boozer%
}{%
Boozer%
}{%
{\protect \APACyear {2005}}%
}]{%
boozer2005physics}
\APACinsertmetastar {%
boozer2005physics}%
\begin{APACrefauthors}%
Boozer, A\BPBI H.%
\end{APACrefauthors}%
\unskip\
\newblock
\APACrefYearMonthDay{2005}{}{}.
\newblock
{\BBOQ}\APACrefatitle {Physics of magnetically confined plasmas} {Physics of
  magnetically confined plasmas}.{\BBCQ}
\newblock
\APACjournalVolNumPages{Reviews of modern physics}{76}{4}{1071}.
\PrintBackRefs{\CurrentBib}

\bibitem [\protect \citeauthoryear {%
Briard%
\ \BBA {} Gomez%
}{%
Briard%
\ \BBA {} Gomez%
}{%
{\protect \APACyear {2018}}%
}]{%
briard2018decay}
\APACinsertmetastar {%
briard2018decay}%
\begin{APACrefauthors}%
Briard, A.%
\BCBT {}\ \BBA {} Gomez, T.%
\end{APACrefauthors}%
\unskip\
\newblock
\APACrefYearMonthDay{2018}{}{}.
\newblock
{\BBOQ}\APACrefatitle {The decay of isotropic magnetohydrodynamics turbulence
  and the effects of cross-helicity} {The decay of isotropic
  magnetohydrodynamics turbulence and the effects of cross-helicity}.{\BBCQ}
\newblock
\APACjournalVolNumPages{Journal of Plasma Physics}{84}{1}{}.
\PrintBackRefs{\CurrentBib}

\bibitem [\protect \citeauthoryear {%
Candelaresi%
\ \BBA {} Del~Sordo%
}{%
Candelaresi%
\ \BBA {} Del~Sordo%
}{%
{\protect \APACyear {2021}}%
}]{%
candelaresi2021stability}
\APACinsertmetastar {%
candelaresi2021stability}%
\begin{APACrefauthors}%
Candelaresi, S.%
\BCBT {}\ \BBA {} Del~Sordo, F.%
\end{APACrefauthors}%
\unskip\
\newblock
\APACrefYearMonthDay{2021}{}{}.
\newblock
{\BBOQ}\APACrefatitle {Stability of plasmas through magnetic helicity}
  {Stability of plasmas through magnetic helicity}.{\BBCQ}
\newblock
\APACjournalVolNumPages{arXiv preprint arXiv:2112.01193}{}{}{}.
\PrintBackRefs{\CurrentBib}

\bibitem [\protect \citeauthoryear {%
Chandran%
}{%
Chandran%
}{%
{\protect \APACyear {2008}}%
}]{%
Chandran}
\APACinsertmetastar {%
Chandran}%
\begin{APACrefauthors}%
Chandran, B\BPBI D\BPBI G.%
\end{APACrefauthors}%
\unskip\
\newblock
\APACrefYearMonthDay{2008}{}{}.
\newblock
{\BBOQ}\APACrefatitle {Strong Anisotropic MHD Turbulence with Cross Helicity}
  {Strong anisotropic mhd turbulence with cross helicity}.{\BBCQ}
\newblock
\APACjournalVolNumPages{The Astrophysical Journal}{685}{}{646-658}.
\PrintBackRefs{\CurrentBib}

\bibitem [\protect \citeauthoryear {%
Faraco%
\ \BBA {} Lindberg%
}{%
Faraco%
\ \BBA {} Lindberg%
}{%
{\protect \APACyear {2020}}%
}]{%
faraco2020proof}
\APACinsertmetastar {%
faraco2020proof}%
\begin{APACrefauthors}%
Faraco, D.%
\BCBT {}\ \BBA {} Lindberg, S.%
\end{APACrefauthors}%
\unskip\
\newblock
\APACrefYearMonthDay{2020}{}{}.
\newblock
{\BBOQ}\APACrefatitle {Proof of Taylor’s conjecture on magnetic helicity
  conservation} {Proof of taylor’s conjecture on magnetic helicity
  conservation}.{\BBCQ}
\newblock
\APACjournalVolNumPages{Communications in Mathematical
  Physics}{373}{2}{707--738}.
\PrintBackRefs{\CurrentBib}

\bibitem [\protect \citeauthoryear {%
Finn%
\ \BBA {} Antonsen~Jr%
}{%
Finn%
\ \BBA {} Antonsen~Jr%
}{%
{\protect \APACyear {1985}}%
}]{%
finn1985magnetic}
\APACinsertmetastar {%
finn1985magnetic}%
\begin{APACrefauthors}%
Finn, J.%
\BCBT {}\ \BBA {} Antonsen~Jr, T.%
\end{APACrefauthors}%
\unskip\
\newblock
\APACrefYearMonthDay{1985}{}{}.
\newblock
{\BBOQ}\APACrefatitle {Magnetic helicity: What is it and what is it good for}
  {Magnetic helicity: What is it and what is it good for}.{\BBCQ}
\newblock
\APACjournalVolNumPages{Comments on Plasma Physics and Controlled
  Fusion}{9}{3}{111--126}.
\PrintBackRefs{\CurrentBib}

\bibitem [\protect \citeauthoryear {%
Goldreich%
\ \BBA {} Sridhar%
}{%
Goldreich%
\ \BBA {} Sridhar%
}{%
{\protect \APACyear {1995}}%
}]{%
goldreich1995toward}
\APACinsertmetastar {%
goldreich1995toward}%
\begin{APACrefauthors}%
Goldreich, P.%
\BCBT {}\ \BBA {} Sridhar, S.%
\end{APACrefauthors}%
\unskip\
\newblock
\APACrefYearMonthDay{1995}{}{}.
\newblock
{\BBOQ}\APACrefatitle {Toward a theory of interstellar turbulence. 2: Strong
  alfvenic turbulence} {Toward a theory of interstellar turbulence. 2: Strong
  alfvenic turbulence}.{\BBCQ}
\newblock
\APACjournalVolNumPages{The Astrophysical Journal}{438}{}{763--775}.
\PrintBackRefs{\CurrentBib}

\bibitem [\protect \citeauthoryear {%
Hazeltine%
\ \BBA {} Meiss%
}{%
Hazeltine%
\ \BBA {} Meiss%
}{%
{\protect \APACyear {2003}}%
}]{%
hazeltine2003plasma}
\APACinsertmetastar {%
hazeltine2003plasma}%
\begin{APACrefauthors}%
Hazeltine, R\BPBI D.%
\BCBT {}\ \BBA {} Meiss, J\BPBI D.%
\end{APACrefauthors}%
\unskip\
\newblock
\APACrefYear{2003}.
\newblock
\APACrefbtitle {Plasma confinement} {Plasma confinement}.
\newblock
\APACaddressPublisher{}{Courier Corporation}.
\PrintBackRefs{\CurrentBib}

\bibitem [\protect \citeauthoryear {%
Heinonen%
, Diamond%
, Katz%
\BCBL {}\ \BBA {} Ronimo%
}{%
Heinonen%
\ \protect \BOthers {.}}{%
{\protect \APACyear {2021}}%
}]{%
heinonen2021role}
\APACinsertmetastar {%
heinonen2021role}%
\begin{APACrefauthors}%
Heinonen, R.%
, Diamond, P.%
, Katz, M.%
\BCBL {}\ \BBA {} Ronimo, G.%
\end{APACrefauthors}%
\unskip\
\newblock
\APACrefYearMonthDay{2021}{}{}.
\newblock
{\BBOQ}\APACrefatitle {On the role of cross-helicity in $\beta$-plane
  magnetohydrodynamic turbulence} {On the role of cross-helicity in
  $\beta$-plane magnetohydrodynamic turbulence}.{\BBCQ}
\newblock
\APACjournalVolNumPages{arXiv preprint arXiv:2103.08091}{}{}{}.
\PrintBackRefs{\CurrentBib}

\bibitem [\protect \citeauthoryear {%
Iovieno%
\ \protect \BOthers {.}}{%
Iovieno%
\ \protect \BOthers {.}}{%
{\protect \APACyear {2016}}%
}]{%
iovieno2016cross}
\APACinsertmetastar {%
iovieno2016cross}%
\begin{APACrefauthors}%
Iovieno, M.%
, Gallana, L.%
, Fraternale, F.%
, Richardson, J.%
, Opher, M.%
\BCBL {}\ \BBA {} Tordella, D.%
\end{APACrefauthors}%
\unskip\
\newblock
\APACrefYearMonthDay{2016}{}{}.
\newblock
{\BBOQ}\APACrefatitle {Cross and magnetic helicity in the outer heliosphere
  from Voyager 2 observations} {Cross and magnetic helicity in the outer
  heliosphere from voyager 2 observations}.{\BBCQ}
\newblock
\APACjournalVolNumPages{European Journal of
  Mechanics-B/Fluids}{55}{}{394--401}.
\PrintBackRefs{\CurrentBib}

\bibitem [\protect \citeauthoryear {%
Knizhnik%
, Antiochos%
, Klimchuk%
\BCBL {}\ \BBA {} DeVore%
}{%
Knizhnik%
\ \protect \BOthers {.}}{%
{\protect \APACyear {2019}}%
}]{%
knizhnik2019role}
\APACinsertmetastar {%
knizhnik2019role}%
\begin{APACrefauthors}%
Knizhnik, K\BPBI J.%
, Antiochos, S\BPBI K.%
, Klimchuk, J\BPBI A.%
\BCBL {}\ \BBA {} DeVore, C\BPBI R.%
\end{APACrefauthors}%
\unskip\
\newblock
\APACrefYearMonthDay{2019}{}{}.
\newblock
{\BBOQ}\APACrefatitle {The role of magnetic helicity in coronal heating} {The
  role of magnetic helicity in coronal heating}.{\BBCQ}
\newblock
\APACjournalVolNumPages{The Astrophysical Journal}{883}{1}{26}.
\PrintBackRefs{\CurrentBib}

\bibitem [\protect \citeauthoryear {%
Kundu%
, Cohen%
\BCBL {}\ \BBA {} Dowling%
}{%
Kundu%
\ \protect \BOthers {.}}{%
{\protect \APACyear {2015}}%
}]{%
kundu2015fluid}
\APACinsertmetastar {%
kundu2015fluid}%
\begin{APACrefauthors}%
Kundu, P\BPBI K.%
, Cohen, I\BPBI M.%
\BCBL {}\ \BBA {} Dowling, D\BPBI R.%
\end{APACrefauthors}%
\unskip\
\newblock
\APACrefYear{2015}.
\newblock
\APACrefbtitle {Fluid mechanics} {Fluid mechanics}.
\newblock
\APACaddressPublisher{}{Academic press}.
\PrintBackRefs{\CurrentBib}

\bibitem [\protect \citeauthoryear {%
Landau%
\ \BBA {} Lifshitz%
}{%
Landau%
\ \BBA {} Lifshitz%
}{%
{\protect \APACyear {1987}}%
}]{%
LANDAU1987192}
\APACinsertmetastar {%
LANDAU1987192}%
\begin{APACrefauthors}%
Landau, L.%
\BCBT {}\ \BBA {} Lifshitz, E.%
\end{APACrefauthors}%
\unskip\
\newblock
\APACrefYearMonthDay{1987}{}{}.
\newblock
{\BBOQ}\APACrefatitle {CHAPTER V - THERMAL CONDUCTION IN FLUIDS} {Chapter v -
  thermal conduction in fluids}.{\BBCQ}
\newblock
\BIn{} L.~Landau\ \BBA {} E.~Lifshitz\ (\BEDS), \APACrefbtitle {Fluid Mechanics
  (Second Edition)} {Fluid mechanics (second edition)}\ (\PrintOrdinal{Second
  Edition}\ \BEd, \BPG~192-226).
\newblock
\APACaddressPublisher{}{Pergamon}.
\newblock
\begin{APACrefURL}
  \url{https://www.sciencedirect.com/science/article/pii/B9780080339337500131}
  \end{APACrefURL}
\newblock
\begin{APACrefDOI} \doi{https://doi.org/10.1016/B978-0-08-033933-7.50013-1}
  \end{APACrefDOI}
\PrintBackRefs{\CurrentBib}

\bibitem [\protect \citeauthoryear {%
Mizeva%
, Stepanov%
\BCBL {}\ \BBA {} Frik%
}{%
Mizeva%
\ \protect \BOthers {.}}{%
{\protect \APACyear {2009}}%
}]{%
mizeva2009cross}
\APACinsertmetastar {%
mizeva2009cross}%
\begin{APACrefauthors}%
Mizeva, I.%
, Stepanov, R.%
\BCBL {}\ \BBA {} Frik, P.%
\end{APACrefauthors}%
\unskip\
\newblock
\APACrefYearMonthDay{2009}{}{}.
\newblock
{\BBOQ}\APACrefatitle {The cross-helicity effect on cascade processes in MHD
  turbulence} {The cross-helicity effect on cascade processes in mhd
  turbulence}.{\BBCQ}
\newblock
\BIn{} \APACrefbtitle {Doklady Physics} {Doklady physics}\ (\BVOL~54, \BPGS\
  93--97).
\PrintBackRefs{\CurrentBib}

\bibitem [\protect \citeauthoryear {%
Mobbs%
}{%
Mobbs%
}{%
{\protect \APACyear {1981}}%
}]{%
mobbs1981some}
\APACinsertmetastar {%
mobbs1981some}%
\begin{APACrefauthors}%
Mobbs, S.%
\end{APACrefauthors}%
\unskip\
\newblock
\APACrefYearMonthDay{1981}{}{}.
\newblock
{\BBOQ}\APACrefatitle {Some vorticity theorems and conservation laws for
  non-barotropic fluids} {Some vorticity theorems and conservation laws for
  non-barotropic fluids}.{\BBCQ}
\newblock
\APACjournalVolNumPages{Journal of Fluid Mechanics}{108}{}{475--483}.
\PrintBackRefs{\CurrentBib}

\bibitem [\protect \citeauthoryear {%
H.~Moffatt%
}{%
H.~Moffatt%
}{%
{\protect \APACyear {1992}}%
}]{%
moffatt1992relaxation}
\APACinsertmetastar {%
moffatt1992relaxation}%
\begin{APACrefauthors}%
Moffatt, H.%
\end{APACrefauthors}%
\unskip\
\newblock
\APACrefYearMonthDay{1992}{}{}.
\newblock
{\BBOQ}\APACrefatitle {Relaxation under topological constraints} {Relaxation
  under topological constraints}.{\BBCQ}
\newblock
\BIn{} \APACrefbtitle {Topological Aspects of the Dynamics of Fluids and
  Plasmas} {Topological aspects of the dynamics of fluids and plasmas}\ (\BPGS\
  3--28).
\newblock
\APACaddressPublisher{}{Springer}.
\PrintBackRefs{\CurrentBib}

\bibitem [\protect \citeauthoryear {%
H\BPBI K.~Moffatt%
}{%
H\BPBI K.~Moffatt%
}{%
{\protect \APACyear {1969}}%
}]{%
moffatt1969degree}
\APACinsertmetastar {%
moffatt1969degree}%
\begin{APACrefauthors}%
Moffatt, H\BPBI K.%
\end{APACrefauthors}%
\unskip\
\newblock
\APACrefYearMonthDay{1969}{}{}.
\newblock
{\BBOQ}\APACrefatitle {The degree of knottedness of tangled vortex lines} {The
  degree of knottedness of tangled vortex lines}.{\BBCQ}
\newblock
\APACjournalVolNumPages{Journal of Fluid Mechanics}{35}{1}{117--129}.
\PrintBackRefs{\CurrentBib}

\bibitem [\protect \citeauthoryear {%
H\BPBI K.~Moffatt%
}{%
H\BPBI K.~Moffatt%
}{%
{\protect \APACyear {1978}}%
}]{%
moffatt1978field}
\APACinsertmetastar {%
moffatt1978field}%
\begin{APACrefauthors}%
Moffatt, H\BPBI K.%
\end{APACrefauthors}%
\unskip\
\newblock
\APACrefYearMonthDay{1978}{}{}.
\newblock
{\BBOQ}\APACrefatitle {Field generation in electrically conducting fluids}
  {Field generation in electrically conducting fluids}.{\BBCQ}
\newblock
\APACjournalVolNumPages{Cambridge University Press, Cambridge, London, New
  York, Melbourne}{2}{}{5--1}.
\PrintBackRefs{\CurrentBib}

\bibitem [\protect \citeauthoryear {%
H\BPBI K.~Moffatt%
\ \BBA {} Ricca%
}{%
H\BPBI K.~Moffatt%
\ \BBA {} Ricca%
}{%
{\protect \APACyear {1995}}%
}]{%
moffatt1995helicity}
\APACinsertmetastar {%
moffatt1995helicity}%
\begin{APACrefauthors}%
Moffatt, H\BPBI K.%
\BCBT {}\ \BBA {} Ricca, R\BPBI L.%
\end{APACrefauthors}%
\unskip\
\newblock
\APACrefYearMonthDay{1995}{}{}.
\newblock
{\BBOQ}\APACrefatitle {Helicity and the C{\u{a}}lug{\u{a}}reanu invariant}
  {Helicity and the c{\u{a}}lug{\u{a}}reanu invariant}.{\BBCQ}
\newblock
\BIn{} \APACrefbtitle {Knots And Applications} {Knots and applications}\
  (\BPGS\ 251--269).
\newblock
\APACaddressPublisher{}{World Scientific}.
\PrintBackRefs{\CurrentBib}

\bibitem [\protect \citeauthoryear {%
Ogilvie%
}{%
Ogilvie%
}{%
{\protect \APACyear {2016}}%
}]{%
ogilvie2016astrophysical}
\APACinsertmetastar {%
ogilvie2016astrophysical}%
\begin{APACrefauthors}%
Ogilvie, G\BPBI I.%
\end{APACrefauthors}%
\unskip\
\newblock
\APACrefYearMonthDay{2016}{}{}.
\newblock
{\BBOQ}\APACrefatitle {Astrophysical fluid dynamics} {Astrophysical fluid
  dynamics}.{\BBCQ}
\newblock
\APACjournalVolNumPages{Journal of Plasma Physics}{82}{3}{}.
\PrintBackRefs{\CurrentBib}

\bibitem [\protect \citeauthoryear {%
Perez%
\ \BBA {} Boldyrev%
}{%
Perez%
\ \BBA {} Boldyrev%
}{%
{\protect \APACyear {2009}}%
}]{%
perez2009role}
\APACinsertmetastar {%
perez2009role}%
\begin{APACrefauthors}%
Perez, J\BPBI C.%
\BCBT {}\ \BBA {} Boldyrev, S.%
\end{APACrefauthors}%
\unskip\
\newblock
\APACrefYearMonthDay{2009}{}{}.
\newblock
{\BBOQ}\APACrefatitle {Role of cross-helicity in magnetohydrodynamic
  turbulence} {Role of cross-helicity in magnetohydrodynamic
  turbulence}.{\BBCQ}
\newblock
\APACjournalVolNumPages{Physical review letters}{102}{2}{025003}.
\PrintBackRefs{\CurrentBib}

\bibitem [\protect \citeauthoryear {%
Priest%
}{%
Priest%
}{%
{\protect \APACyear {2014}}%
}]{%
priest2014magnetohydrodynamics}
\APACinsertmetastar {%
priest2014magnetohydrodynamics}%
\begin{APACrefauthors}%
Priest, E.%
\end{APACrefauthors}%
\unskip\
\newblock
\APACrefYear{2014}.
\newblock
\APACrefbtitle {Magnetohydrodynamics of the Sun} {Magnetohydrodynamics of the
  sun}.
\newblock
\APACaddressPublisher{}{Cambridge University Press}.
\PrintBackRefs{\CurrentBib}

\bibitem [\protect \citeauthoryear {%
Sakurai%
}{%
Sakurai%
}{%
{\protect \APACyear {1979}}%
}]{%
sakurai1979new}
\APACinsertmetastar {%
sakurai1979new}%
\begin{APACrefauthors}%
Sakurai, T.%
\end{APACrefauthors}%
\unskip\
\newblock
\APACrefYearMonthDay{1979}{}{}.
\newblock
{\BBOQ}\APACrefatitle {A New Approach to the Force-Free Field and Its
  Application to the Mag-netic Field of Solar Active Regions} {A new approach
  to the force-free field and its application to the mag-netic field of solar
  active regions}.{\BBCQ}
\newblock
\APACjournalVolNumPages{Publications of the Astronomical Society of
  Japan}{31}{}{209--230}.
\PrintBackRefs{\CurrentBib}

\bibitem [\protect \citeauthoryear {%
Smith%
, Stawarz%
, Vasquez%
, Forman%
\BCBL {}\ \BBA {} MacBride%
}{%
Smith%
\ \protect \BOthers {.}}{%
{\protect \APACyear {2009}}%
}]{%
smith2009turbulent}
\APACinsertmetastar {%
smith2009turbulent}%
\begin{APACrefauthors}%
Smith, C\BPBI W.%
, Stawarz, J\BPBI E.%
, Vasquez, B\BPBI J.%
, Forman, M\BPBI A.%
\BCBL {}\ \BBA {} MacBride, B\BPBI T.%
\end{APACrefauthors}%
\unskip\
\newblock
\APACrefYearMonthDay{2009}{}{}.
\newblock
{\BBOQ}\APACrefatitle {Turbulent cascade at 1 AU in high cross-helicity flows}
  {Turbulent cascade at 1 au in high cross-helicity flows}.{\BBCQ}
\newblock
\APACjournalVolNumPages{Physical review letters}{103}{20}{201101}.
\PrintBackRefs{\CurrentBib}

\bibitem [\protect \citeauthoryear {%
Sturrock%
}{%
Sturrock%
}{%
{\protect \APACyear {1994}}%
}]{%
sturrock1994plasma}
\APACinsertmetastar {%
sturrock1994plasma}%
\begin{APACrefauthors}%
Sturrock, P\BPBI A.%
\end{APACrefauthors}%
\unskip\
\newblock
\APACrefYear{1994}.
\newblock
\APACrefbtitle {Plasma physics: an introduction to the theory of astrophysical,
  geophysical and laboratory plasmas} {Plasma physics: an introduction to the
  theory of astrophysical, geophysical and laboratory plasmas}.
\newblock
\APACaddressPublisher{}{Cambridge University Press}.
\PrintBackRefs{\CurrentBib}

\bibitem [\protect \citeauthoryear {%
M.~Verma%
, Sharma%
, Chatterjee%
\BCBL {}\ \BBA {} Alam%
}{%
M.~Verma%
\ \protect \BOthers {.}}{%
{\protect \APACyear {2021}}%
}]{%
verma2021variable}
\APACinsertmetastar {%
verma2021variable}%
\begin{APACrefauthors}%
Verma, M.%
, Sharma, M.%
, Chatterjee, S.%
\BCBL {}\ \BBA {} Alam, S.%
\end{APACrefauthors}%
\unskip\
\newblock
\APACrefYearMonthDay{2021}{}{}.
\newblock
{\BBOQ}\APACrefatitle {Variable energy fluxes and exact relations in
  Magnetohydrodynamics turbulence} {Variable energy fluxes and exact relations
  in magnetohydrodynamics turbulence}.{\BBCQ}
\newblock
\APACjournalVolNumPages{Fluids}{6}{6}{225}.
\PrintBackRefs{\CurrentBib}

\bibitem [\protect \citeauthoryear {%
M\BPBI K.~Verma%
}{%
M\BPBI K.~Verma%
}{%
{\protect \APACyear {2004}}%
}]{%
verma2004statistical}
\APACinsertmetastar {%
verma2004statistical}%
\begin{APACrefauthors}%
Verma, M\BPBI K.%
\end{APACrefauthors}%
\unskip\
\newblock
\APACrefYearMonthDay{2004}{}{}.
\newblock
{\BBOQ}\APACrefatitle {Statistical theory of magnetohydrodynamic turbulence:
  recent results} {Statistical theory of magnetohydrodynamic turbulence: recent
  results}.{\BBCQ}
\newblock
\APACjournalVolNumPages{Physics Reports}{401}{5-6}{229--380}.
\PrintBackRefs{\CurrentBib}

\bibitem [\protect \citeauthoryear {%
M\BPBI K.~Verma%
}{%
M\BPBI K.~Verma%
}{%
{\protect \APACyear {2019}}%
}]{%
verma2019energy}
\APACinsertmetastar {%
verma2019energy}%
\begin{APACrefauthors}%
Verma, M\BPBI K.%
\end{APACrefauthors}%
\unskip\
\newblock
\APACrefYear{2019}.
\newblock
\APACrefbtitle {Energy transfers in fluid flows: multiscale and spectral
  perspectives} {Energy transfers in fluid flows: multiscale and spectral
  perspectives}.
\newblock
\APACaddressPublisher{}{Cambridge University Press}.
\PrintBackRefs{\CurrentBib}

\bibitem [\protect \citeauthoryear {%
M\BPBI K.~Verma%
}{%
M\BPBI K.~Verma%
}{%
{\protect \APACyear {2021}}%
}]{%
Verma_2021}
\APACinsertmetastar {%
Verma_2021}%
\begin{APACrefauthors}%
Verma, M\BPBI K.%
\end{APACrefauthors}%
\unskip\
\newblock
\APACrefYearMonthDay{2021}{dec}{}.
\newblock
{\BBOQ}\APACrefatitle {Variable energy flux in turbulence} {Variable energy
  flux in turbulence}.{\BBCQ}
\newblock
\APACjournalVolNumPages{Journal of Physics A: Mathematical and
  Theoretical}{55}{1}{013002}.
\newblock
\begin{APACrefURL} \url{https://doi.org/10.1088/1751-8121/ac354e}
  \end{APACrefURL}
\newblock
\begin{APACrefDOI} \doi{10.1088/1751-8121/ac354e} \end{APACrefDOI}
\PrintBackRefs{\CurrentBib}

\bibitem [\protect \citeauthoryear {%
Webb%
\ \BBA {} Anco%
}{%
Webb%
\ \BBA {} Anco%
}{%
{\protect \APACyear {2017}}%
}]{%
webb2017magnetohydrodynamic}
\APACinsertmetastar {%
webb2017magnetohydrodynamic}%
\begin{APACrefauthors}%
Webb, G.%
\BCBT {}\ \BBA {} Anco, S.%
\end{APACrefauthors}%
\unskip\
\newblock
\APACrefYearMonthDay{2017}{}{}.
\newblock
{\BBOQ}\APACrefatitle {On magnetohydrodynamic gauge field theory} {On
  magnetohydrodynamic gauge field theory}.{\BBCQ}
\newblock
\APACjournalVolNumPages{Journal of Physics A: Mathematical and
  Theoretical}{50}{25}{255501}.
\PrintBackRefs{\CurrentBib}

\bibitem [\protect \citeauthoryear {%
Webb%
, Dasgupta%
, McKenzie%
, Hu%
\BCBL {}\ \BBA {} Zank%
}{%
Webb%
\ \protect \BOthers {.}}{%
{\protect \APACyear {2014}}%
{\protect \APACexlab {{\protect \BCnt {1}}}}}]{%
webb2014localBB}
\APACinsertmetastar {%
webb2014localBB}%
\begin{APACrefauthors}%
Webb, G.%
, Dasgupta, B.%
, McKenzie, J.%
, Hu, Q.%
\BCBL {}\ \BBA {} Zank, G.%
\end{APACrefauthors}%
\unskip\
\newblock
\APACrefYearMonthDay{2014{\protect \BCnt {1}}}{}{}.
\newblock
{\BBOQ}\APACrefatitle {Local and nonlocal advected invariants and helicities in
  magnetohydrodynamics and gas dynamics: II. Noether's theorems and Casimirs}
  {Local and nonlocal advected invariants and helicities in
  magnetohydrodynamics and gas dynamics: Ii. noether's theorems and
  casimirs}.{\BBCQ}
\newblock
\APACjournalVolNumPages{Journal of Physics A: Mathematical and
  Theoretical}{47}{9}{095502}.
\PrintBackRefs{\CurrentBib}

\bibitem [\protect \citeauthoryear {%
Webb%
, Dasgupta%
, McKenzie%
, Hu%
\BCBL {}\ \BBA {} Zank%
}{%
Webb%
\ \protect \BOthers {.}}{%
{\protect \APACyear {2014}}%
{\protect \APACexlab {{\protect \BCnt {2}}}}}]{%
webb2014localAA}
\APACinsertmetastar {%
webb2014localAA}%
\begin{APACrefauthors}%
Webb, G.%
, Dasgupta, B.%
, McKenzie, J.%
, Hu, Q.%
\BCBL {}\ \BBA {} Zank, G.%
\end{APACrefauthors}%
\unskip\
\newblock
\APACrefYearMonthDay{2014{\protect \BCnt {2}}}{}{}.
\newblock
{\BBOQ}\APACrefatitle {Local and nonlocal advected invariants and helicities in
  magnetohydrodynamics and gas dynamics I: Lie dragging approach} {Local and
  nonlocal advected invariants and helicities in magnetohydrodynamics and gas
  dynamics i: Lie dragging approach}.{\BBCQ}
\newblock
\APACjournalVolNumPages{Journal of Physics A: Mathematical and
  Theoretical}{47}{9}{095501}.
\PrintBackRefs{\CurrentBib}

\bibitem [\protect \citeauthoryear {%
Webb%
\ \BBA {} Mace%
}{%
Webb%
\ \BBA {} Mace%
}{%
{\protect \APACyear {2015}}%
}]{%
webb2015potential}
\APACinsertmetastar {%
webb2015potential}%
\begin{APACrefauthors}%
Webb, G.%
\BCBT {}\ \BBA {} Mace, R.%
\end{APACrefauthors}%
\unskip\
\newblock
\APACrefYearMonthDay{2015}{}{}.
\newblock
{\BBOQ}\APACrefatitle {Potential vorticity in magnetohydrodynamics} {Potential
  vorticity in magnetohydrodynamics}.{\BBCQ}
\newblock
\APACjournalVolNumPages{Journal of Plasma Physics}{81}{1}{}.
\PrintBackRefs{\CurrentBib}

\bibitem [\protect \citeauthoryear {%
Webb%
, McKenzie%
\BCBL {}\ \BBA {} Zank%
}{%
Webb%
\ \protect \BOthers {.}}{%
{\protect \APACyear {2015}}%
}]{%
webb2015multi}
\APACinsertmetastar {%
webb2015multi}%
\begin{APACrefauthors}%
Webb, G.%
, McKenzie, J.%
\BCBL {}\ \BBA {} Zank, G.%
\end{APACrefauthors}%
\unskip\
\newblock
\APACrefYearMonthDay{2015}{}{}.
\newblock
{\BBOQ}\APACrefatitle {Multi-symplectic magnetohydrodynamics: II, addendum and
  erratum} {Multi-symplectic magnetohydrodynamics: Ii, addendum and
  erratum}.{\BBCQ}
\newblock
\APACjournalVolNumPages{Journal of Plasma Physics}{81}{6}{}.
\PrintBackRefs{\CurrentBib}

\bibitem [\protect \citeauthoryear {%
Woltjer%
}{%
Woltjer%
}{%
{\protect \APACyear {1958}}%
{\protect \APACexlab {{\protect \BCnt {1}}}}}]{%
woltjer1958hydromagnetic}
\APACinsertmetastar {%
woltjer1958hydromagnetic}%
\begin{APACrefauthors}%
Woltjer, L.%
\end{APACrefauthors}%
\unskip\
\newblock
\APACrefYearMonthDay{1958{\protect \BCnt {1}}}{}{}.
\newblock
{\BBOQ}\APACrefatitle {On hydromagnetic equilibrium} {On hydromagnetic
  equilibrium}.{\BBCQ}
\newblock
\APACjournalVolNumPages{Proceedings of the National Academy of Sciences of the
  United States of America}{44}{9}{833}.
\PrintBackRefs{\CurrentBib}

\bibitem [\protect \citeauthoryear {%
Woltjer%
}{%
Woltjer%
}{%
{\protect \APACyear {1958}}%
{\protect \APACexlab {{\protect \BCnt {2}}}}}]{%
woltjer1958theorem}
\APACinsertmetastar {%
woltjer1958theorem}%
\begin{APACrefauthors}%
Woltjer, L.%
\end{APACrefauthors}%
\unskip\
\newblock
\APACrefYearMonthDay{1958{\protect \BCnt {2}}}{}{}.
\newblock
{\BBOQ}\APACrefatitle {A theorem on force-free magnetic fields} {A theorem on
  force-free magnetic fields}.{\BBCQ}
\newblock
\APACjournalVolNumPages{Proceedings of the National Academy of Sciences of the
  United States of America}{44}{6}{489}.
\PrintBackRefs{\CurrentBib}

\bibitem [\protect \citeauthoryear {%
Yahalom%
}{%
Yahalom%
}{%
{\protect \APACyear {1995}}%
}]{%
yahalom1995helicity}
\APACinsertmetastar {%
yahalom1995helicity}%
\begin{APACrefauthors}%
Yahalom, A.%
\end{APACrefauthors}%
\unskip\
\newblock
\APACrefYearMonthDay{1995}{}{}.
\newblock
{\BBOQ}\APACrefatitle {Helicity conservation via the Noether theorem} {Helicity
  conservation via the noether theorem}.{\BBCQ}
\newblock
\APACjournalVolNumPages{Journal of Mathematical Physics}{36}{3}{1324--1327}.
\PrintBackRefs{\CurrentBib}

\bibitem [\protect \citeauthoryear {%
Yahalom%
}{%
Yahalom%
}{%
{\protect \APACyear {2003}}%
}]{%
yahalom2003method}
\APACinsertmetastar {%
yahalom2003method}%
\begin{APACrefauthors}%
Yahalom, A.%
\end{APACrefauthors}%
\unskip\
\newblock
\APACrefYearMonthDay{2003}{{\APACmonth{02}}~4}{}.
\newblock
\APACrefbtitle {Method and system for numerical simulation of fluid flow.}
  {Method and system for numerical simulation of fluid flow.}
\newblock
\APACaddressPublisher{}{Google Patents}.
\newblock
\APACrefnote{US Patent 6,516,292}
\PrintBackRefs{\CurrentBib}

\bibitem [\protect \citeauthoryear {%
Yahalom%
}{%
Yahalom%
}{%
{\protect \APACyear {2013}}%
}]{%
yahalom2013aharonov}
\APACinsertmetastar {%
yahalom2013aharonov}%
\begin{APACrefauthors}%
Yahalom, A.%
\end{APACrefauthors}%
\unskip\
\newblock
\APACrefYearMonthDay{2013}{}{}.
\newblock
{\BBOQ}\APACrefatitle {Aharonov--Bohm effects in magnetohydrodynamics}
  {Aharonov--bohm effects in magnetohydrodynamics}.{\BBCQ}
\newblock
\APACjournalVolNumPages{Physics Letters A}{377}{31-33}{1898--1904}.
\PrintBackRefs{\CurrentBib}

\bibitem [\protect \citeauthoryear {%
Yahalom%
}{%
Yahalom%
}{%
{\protect \APACyear {2016}}%
}]{%
yahalom2016simplified}
\APACinsertmetastar {%
yahalom2016simplified}%
\begin{APACrefauthors}%
Yahalom, A.%
\end{APACrefauthors}%
\unskip\
\newblock
\APACrefYearMonthDay{2016}{}{}.
\newblock
{\BBOQ}\APACrefatitle {Simplified variational principles for non-barotropic
  magnetohydrodynamics} {Simplified variational principles for non-barotropic
  magnetohydrodynamics}.{\BBCQ}
\newblock
\APACjournalVolNumPages{Journal of Plasma Physics}{82}{2}{}.
\PrintBackRefs{\CurrentBib}

\bibitem [\protect \citeauthoryear {%
Yahalom%
}{%
Yahalom%
}{%
{\protect \APACyear {2017}}%
{\protect \APACexlab {{\protect \BCnt {1}}}}}]{%
yahalom2017conserved}
\APACinsertmetastar {%
yahalom2017conserved}%
\begin{APACrefauthors}%
Yahalom, A.%
\end{APACrefauthors}%
\unskip\
\newblock
\APACrefYearMonthDay{2017{\protect \BCnt {1}}}{}{}.
\newblock
{\BBOQ}\APACrefatitle {A conserved local cross helicity for non-barotropic MHD}
  {A conserved local cross helicity for non-barotropic mhd}.{\BBCQ}
\newblock
\APACjournalVolNumPages{Geophysical \& Astrophysical Fluid
  Dynamics}{111}{2}{131--137}.
\PrintBackRefs{\CurrentBib}

\bibitem [\protect \citeauthoryear {%
Yahalom%
}{%
Yahalom%
}{%
{\protect \APACyear {2017}}%
{\protect \APACexlab {{\protect \BCnt {2}}}}}]{%
yahalom2017non}
\APACinsertmetastar {%
yahalom2017non}%
\begin{APACrefauthors}%
Yahalom, A.%
\end{APACrefauthors}%
\unskip\
\newblock
\APACrefYearMonthDay{2017{\protect \BCnt {2}}}{}{}.
\newblock
{\BBOQ}\APACrefatitle {Non-barotropic cross-helicity conservation applications
  in magnetohydrodynamics and the Aharanov--Bohm effect} {Non-barotropic
  cross-helicity conservation applications in magnetohydrodynamics and the
  aharanov--bohm effect}.{\BBCQ}
\newblock
\APACjournalVolNumPages{Fluid Dynamics Research}{50}{1}{011406}.
\PrintBackRefs{\CurrentBib}

\bibitem [\protect \citeauthoryear {%
Yahalom%
}{%
Yahalom%
}{%
{\protect \APACyear {2019}}%
}]{%
yahalom2019new}
\APACinsertmetastar {%
yahalom2019new}%
\begin{APACrefauthors}%
Yahalom, A.%
\end{APACrefauthors}%
\unskip\
\newblock
\APACrefYearMonthDay{2019}{}{}.
\newblock
{\BBOQ}\APACrefatitle {A new diffeomorphism symmetry group of non-barotropic
  magnetohydrodynamics} {A new diffeomorphism symmetry group of non-barotropic
  magnetohydrodynamics}.{\BBCQ}
\newblock
\BIn{} \APACrefbtitle {Journal of Physics: Conference Series} {Journal of
  physics: Conference series}\ (\BVOL\ 1194, \BPG~012113).
\PrintBackRefs{\CurrentBib}

\bibitem [\protect \citeauthoryear {%
Yahalom%
\ \BBA {} Lynden-Bell%
}{%
Yahalom%
\ \BBA {} Lynden-Bell%
}{%
{\protect \APACyear {2008}}%
}]{%
yahalom2008simplified}
\APACinsertmetastar {%
yahalom2008simplified}%
\begin{APACrefauthors}%
Yahalom, A.%
\BCBT {}\ \BBA {} Lynden-Bell, D.%
\end{APACrefauthors}%
\unskip\
\newblock
\APACrefYearMonthDay{2008}{}{}.
\newblock
{\BBOQ}\APACrefatitle {Simplified variational principles for barotropic
  magnetohydrodynamics} {Simplified variational principles for barotropic
  magnetohydrodynamics}.{\BBCQ}
\newblock
\APACjournalVolNumPages{Journal of Fluid Mechanics}{607}{}{235}.
\PrintBackRefs{\CurrentBib}

\bibitem [\protect \citeauthoryear {%
Yahalom%
\ \BBA {} Qin%
}{%
Yahalom%
\ \BBA {} Qin%
}{%
{\protect \APACyear {2021}}%
}]{%
yahalom2021noether}
\APACinsertmetastar {%
yahalom2021noether}%
\begin{APACrefauthors}%
Yahalom, A.%
\BCBT {}\ \BBA {} Qin, H.%
\end{APACrefauthors}%
\unskip\
\newblock
\APACrefYearMonthDay{2021}{}{}.
\newblock
{\BBOQ}\APACrefatitle {Noether currents for Eulerian variational principles in
  non-barotropic magnetohydrodynamics and topological conservations laws}
  {Noether currents for eulerian variational principles in non-barotropic
  magnetohydrodynamics and topological conservations laws}.{\BBCQ}
\newblock
\APACjournalVolNumPages{Journal of Fluid Mechanics}{908}{}{}.
\PrintBackRefs{\CurrentBib}

\bibitem [\protect \citeauthoryear {%
Yokoi%
}{%
Yokoi%
}{%
{\protect \APACyear {2013}}%
}]{%
yokoi2013cross}
\APACinsertmetastar {%
yokoi2013cross}%
\begin{APACrefauthors}%
Yokoi, N.%
\end{APACrefauthors}%
\unskip\
\newblock
\APACrefYearMonthDay{2013}{}{}.
\newblock
{\BBOQ}\APACrefatitle {Cross helicity and related dynamo} {Cross helicity and
  related dynamo}.{\BBCQ}
\newblock
\APACjournalVolNumPages{Geophysical \& Astrophysical Fluid
  Dynamics}{107}{1-2}{114--184}.
\PrintBackRefs{\CurrentBib}

\bibitem [\protect \citeauthoryear {%
Zank%
\ \protect \BOthers {.}}{%
Zank%
\ \protect \BOthers {.}}{%
{\protect \APACyear {2011}}%
}]{%
zank2011transport}
\APACinsertmetastar {%
zank2011transport}%
\begin{APACrefauthors}%
Zank, G.%
, Dosch, A.%
, Hunana, P.%
, Florinski, V.%
, Matthaeus, W.%
\BCBL {}\ \BBA {} Webb, G.%
\end{APACrefauthors}%
\unskip\
\newblock
\APACrefYearMonthDay{2011}{}{}.
\newblock
{\BBOQ}\APACrefatitle {The transport of low-frequency turbulence in
  astrophysical flows. I. Governing equations} {The transport of low-frequency
  turbulence in astrophysical flows. i. governing equations}.{\BBCQ}
\newblock
\APACjournalVolNumPages{The Astrophysical Journal}{745}{1}{35}.
\PrintBackRefs{\CurrentBib}

\bibitem [\protect \citeauthoryear {%
Zhou%
\ \BBA {} Matthaeus%
}{%
Zhou%
\ \BBA {} Matthaeus%
}{%
{\protect \APACyear {1990}}%
{\protect \APACexlab {{\protect \BCnt {1}}}}}]{%
zhou1990models}
\APACinsertmetastar {%
zhou1990models}%
\begin{APACrefauthors}%
Zhou, Y.%
\BCBT {}\ \BBA {} Matthaeus, W\BPBI H.%
\end{APACrefauthors}%
\unskip\
\newblock
\APACrefYearMonthDay{1990{\protect \BCnt {1}}}{}{}.
\newblock
{\BBOQ}\APACrefatitle {Models of inertial range spectra of interplanetary
  magnetohydrodynamic turbulence} {Models of inertial range spectra of
  interplanetary magnetohydrodynamic turbulence}.{\BBCQ}
\newblock
\APACjournalVolNumPages{Journal of Geophysical Research: Space
  Physics}{95}{A9}{14881--14892}.
\PrintBackRefs{\CurrentBib}

\bibitem [\protect \citeauthoryear {%
Zhou%
\ \BBA {} Matthaeus%
}{%
Zhou%
\ \BBA {} Matthaeus%
}{%
{\protect \APACyear {1990}}%
{\protect \APACexlab {{\protect \BCnt {2}}}}}]{%
zhou1990transport}
\APACinsertmetastar {%
zhou1990transport}%
\begin{APACrefauthors}%
Zhou, Y.%
\BCBT {}\ \BBA {} Matthaeus, W\BPBI H.%
\end{APACrefauthors}%
\unskip\
\newblock
\APACrefYearMonthDay{1990{\protect \BCnt {2}}}{}{}.
\newblock
{\BBOQ}\APACrefatitle {Transport and turbulence modeling of solar wind
  fluctuations} {Transport and turbulence modeling of solar wind
  fluctuations}.{\BBCQ}
\newblock
\APACjournalVolNumPages{Journal of Geophysical Research: Space
  Physics}{95}{A7}{10291--10311}.
\PrintBackRefs{\CurrentBib}

\end{thebibliography}

\end{document}